\documentclass[a4paper,11pt]{article}
\pdfoutput=1 % if your are submitting a pdflatex (i.e. if you have
\usepackage{jcappub} % for details on the use of the package, please
\usepackage[T1]{fontenc} % if needed

\usepackage{amssymb}
\usepackage{amsmath}
\usepackage{natbib}
\def\simgt{\hbox{\,\rlap{\raise 0.425ex\hbox{$>$}}\lower 0.65ex\hbox{$\sim$}\,}}
\def\simlt{\hbox{\,\rlap{\raise 0.425ex\hbox{$<$}}\lower 0.65ex\hbox{$\sim$}\,}}
\definecolor{darkgreen}{rgb}{0,0.5,0}
\definecolor{mag}{rgb}{0.79,0.08,0.48}
\definecolor{darkblue}{rgb}{0,0.2,0.8}

\title{\boldmath {Testing DARKexp against energy and density distributions of Millennium-II halos}}

\author[a,1]{Chris Nolting,\note{Corresponding author.}}
\author[a]{Liliya L.R. Williams,}
\author[b]{Michael Boylan-Kolchin}
\author[c]{and Jens Hjorth}

\affiliation[a]{School of Physics and Astronomy, University of Minnesota, 116 Church Street SE, Minneapolis, MN 55454, USA}
\affiliation[b]{Department of Astronomy, The University of Texas at Austin, 2515 Speedway, Stop C1400, Austin, TX 78712, USA}
\affiliation[c]{Dark Cosmology Centre, Niels Bohr Institute, University of Copenhagen, Juliane Maries Vej 30, DK-2100 Copenhagen, Denmark}

\emailAdd{nolting@astro.umn.edu}
\emailAdd{llrw@astro.umn.edu}
\emailAdd{mbk@astro.as.utexas.edu}
\emailAdd{jens@dark-cosmology.dk}

\abstract{
We test the DARKexp model for relaxed, self-gravitating, collisionless systems against equilibrium dark matter halos from the Millennium-II simulation. While limited tests of DARKexp against simulations and observations have been carried out elsewhere, this is the first time the testing is done with a large sample of simulated halos spanning a factor of $\sim 50$ in mass, and using independent fits to density and energy distributions. We show that DARKexp, a one shape parameter family, provides very good fits to the shapes of density profiles, $\rho(r)$, and differential energy distributions, $N(E)$, of individual simulated halos. The best fit shape parameter $\phi_0$ obtained from the two types of fits are correlated, though with scatter. Our most important conclusions come from $\rho(r)$ and $N(E)$ that have been averaged over many halos. These show that the bulk of the deviations between DARKexp and individual Millennium-II halos come from halo-to-halo fluctuations, likely driven by substructure, and other density perturbations. The average $\rho(r)$ and $N(E)$ are quite smooth and follow DARKexp very closely. The only deviation that remains after averaging is small, and located at most bound energies for $N(E)$ and smallest radii for $\rho(r)$. Since the deviation is confined to 3-4 smoothing lengths, and is larger for low mass halos, it is likely due to numerical resolution effects. 
}
\begin{document}
\maketitle
\flushbottom

\section{Introduction}

Virialized systems such as galaxies and galaxy clusters are the building blocks of large-scale structure. How these systems relax and how to describe their final steady state are two of the fundamental questions in cosmology.  Both questions are most commonly addressed through high resolution N-body simulations. In particular, the density profile shape of pure dark matter halos is now firmly established \citep{dc91,nav97}, and is well described by empirical Navarro-Frenk-White
\citep[hereafter, NFW]{nav97,nav04} or Einasto \citep{einasto65} fitting formulae.

Much of the vast literature on the formation and evolution of galaxies and galaxy clusters requires modeling the density
profiles of virialized dark matter halos. This includes such diverse applications as: attempts to solve the small scale problems of $\Lambda$CDM \citep{weinberg13};  analysis of the growth of bound structures \citep{bosch14}; predictions of
the dark matter annihilation signal from the Milky Way and other galaxies \citep{lavalle15}; and comparison to profiles of clusters obtained from lensing \citep{umetsu14,newman13}. Most often, such works adopt NFW or Einasto profiles.

Despite their wide utility, the theoretical explanation of the shapes of these profiles is still missing. This unfortunate state of affairs can be compared to that of the studies of stars, in which astronomers have been able to solve for the interior structure of stars in equilibrium for about a century now \citep{edd}, without numerical simulations of a large collection of self-gravitating gas particles. Study of dark matter halos should also be able to theoretically predict the structure of relaxed halos.

Numerous attempts, using a variety of approaches, have been made to explain how NFW or Einasto profiles arise. 
Some of the earliest methods used secondary infall models \citep{gg72,rg87} as the starting point, where an initial overdensity grows by accreting  mass from its surroundings. To implement secondary infall one needs to specify the mass infall rate, or mass accretion history; an early example of this method is \cite{ar98}. The accretion rate can be obtained, for example, from the extended Press-Schechter formalism \citep{man03,gc07}. The main conclusion from these studies is that the radially averaged halo properties are a direct consequence of the mass accretion history. To get a more accurate final profile, \cite{lu06} take the accretion rate from N-body simulations, as quantified by \cite{wech02}. They claim that early fast accretion results in the inner $\rho\sim r^{-1}$ slope if the particle orbits are allowed to ``isotropize'', while the later slower accretion gives the outer $\rho\sim r^{-3}$ slope. A similar study was done by \cite{sh04}, who also used mass accretion rate, but used fluid approximation in the collisionless Boltzmann equation, and hence isotropic velocity dispersion throughout the halo. \cite{asca07} conclude that the secondary infall provides a viable dynamical model for predicting the structure and evolution of the density profile of dark matter halos. 

In other words, this implies that to produce density profiles of dark matter systems it is sufficient to consider isolated collapse, and disregard the hierarchical nature of structure formation seen in N-body simulations and the real universe \citep{wbd04,asca07,wang09}. This important result is implicitly used by later works which also rely on secondary infall, but emphasize additional physical processes as being important. For example, \cite{dalal10} argues that the main ingredients are spherical collapse of density peaks and dynamical friction. \cite{salvador12} relied on inside-out growth to keep the inner structure essentially unchanged, instead of using adiabatic invariants for halo collapse, as was done by several other earlier authors.

One physical process has long been suspected to be involved in shaping the density profile of relaxed halos: radial orbit instability (ROI) \citep{huss99}. \cite{barnes05} and \cite{bellovary08} show that ROI sets the scale length at which the velocity dispersion changes rapidly from isotropic to radially anisotropic, which is related to the scale of the density profile slope change. \cite{visbal12} argue that ROI is the key physical mechanism and that its action can be captured by a simple distribution function, which matches the distribution function of simulated halos with 10 adjustable parameters.

A very different approach to explaining halo profiles is based on the ansatz that the near universal density profiles are the result of a relaxation process whose final state can be obtained as a maximum entropy state. This approach was pioneered by \cite{ogorod57} and \cite{lb67}. The latter coined the phrase `violent relaxation', to refer to the fast process through which galaxies relax. However, the results obtained by these authors have at least two problems: the density profile---the isothermal sphere---has infinite mass and energy, and does not match results of simulations. One possible way to resolve these problems is to postulate that the relaxation that galaxies and dark matter halos undergo is incomplete. For example, \cite {sb87} introduced additional constraints in to their distribution functions, while \cite{hm91} assumed that the relaxation region in configuration space is limited to the central halo.  \cite{wn87} used the entropy of perfect gas and investigated what density profile predictions can be obtained through maximizing entropy, but their results are a poor fit to observations. \cite{hk10} tried a modification of the same method in combination with anisotropic velocity dispersions, with more promising results. 

Another approach of note is the work of \cite{sh92} who proposed a scattering model for violent relaxation, and derive the corresponding differential energy distribution, $dM/dE=N(E)$. \cite{pg13} make use of maximum entropy together with physically motivated dynamical constraints, which are encoded in several parameters, to obtain a phase-space distribution of virialized dark matter halos. 

Despite the apparent inadequacies of the maximum entropy approach pointed out in several papers, it is still an intriguing possibility that density profiles of dark matter halos have their origin in statistical mechanics of self-gravitating systems.  As \cite{ogorod57} and \cite{lb67}, we start with that assumption.  Our derivation, presented in \cite{paperI}, is similar to that in \cite{lb67}, or the classical derivation that leads to the Maxwell-Boltzmann distribution. We partition the system into cells each having occupation number $n_i$ and degeneracy $g_i$, and then write down the number of possible states, $W$, accessible to the system, as a function of $n_i$ and $g_i$. Next, we optimize $\ln W$ under the constraints of fixed total mass and energy. In addition to these steps which are the same as in the classical derivation, we introduce two key modifications. Namely, we discard the Sterling approximation in favor of a more superior one, and interpret the occupation number $n_i$ to be the number of particles in energy space, and not in phase space. With these, we arrive at DARKexp, whose differential energy distribution is given by:
\begin{equation}
N(E)=A[\exp(\beta(\Phi_0 - E))-1]=A[\exp(\phi_0 - \epsilon)-1]\,, \label{DARKexpEnergy}
\end{equation}
where $N(E)\,\Delta E$ is the number of particles with energy per unit mass in the range $E \pm \frac{1}{2}\Delta E$,  $\beta$ is the inverse temperature, $\Phi_0$ is the depth of the central potential, $\epsilon=\beta E$ is the dimensionless energy, and $A$ is a mass normalization constant. $\phi_0=\beta\Phi_0$ is the only shape parameter of the model, and can be traced back to two Lagrange multiplies used to keep total mass and energy fixed during entropy maximization.  $N(E)$ is related to the more familiar phase space density $f(E)$ through the density of states $g(E)$: $N(E)=f(E)\,g(E)$. Note that DARKexp implies complete relaxation in energy. The density profile corresponding to a given value of $\phi_0$ can be found by an iterative procedure \citep{binney82}, and was used in \cite{paperII}. It is interesting to note that \cite{binney82} noticed that the energy distribution of the de Vaucouleurs profile can be approximated by $N(E)=A\exp(-\beta E)$, which is similar to DARKexp, eq.~\ref{DARKexpEnergy}. The differences are confined to the very bound and the least bound energy limits.

While further theoretical work is still needed on DARKexp, one way to test the model is to compare it to observations and simulations. Relaxed, dark matter dominated systems, like relaxed galaxy clusters, are a good test-bed because they have a large dynamic range of observationally constrained density profiles. \cite{silva2013} obtained a stacked profile from four $z\approx 0.3$ clusters and compared it to phenomenological density profiles (NFW, Einasto, Sersic, Stadel, Baltz-Marshall-Oguri [BMO], Hernquist) and theoretical (non-singular Isothermal Sphere, DARKexp and the model of \cite{hk10}) models of the dark matter distribution.  They find that among the theoretical profiles, i.e. those based on some physical argument, the DARKexp model has the best performance, very close to that of the best performing phenomenological BMO profile.  More recently, \citep{umetsu15} use CLASH galaxy clusters to derive an ensemble-averaged surface mass density profile from 16 individual equilibrium clusters. The stacked mass profile is well described by phenomenological NFW and Einasto profiles and theoretical DARKexp model.

Simulated dark matter halos provide a better comparison test, because all six positional and velocity pieces of information are available for each particle, with no projection effects or observational uncertainties. To come up with a fitting formula, like NFW or Einasto, only the density profiles of simulated systems are used. Because DARKexp has its origins in a model of the energy distribution of collisionless particles, the comparison we carry out in this paper includes both density profiles and energy distributions. Partial testing against a set of simulated clusters was done in \cite{paperIII}, \cite{paperIV} and \cite{hjorth15} using halos from the Bolshoi simulation \citep{bolshoi}. However, these papers used either density or energy distributions, not both, and some were limited to a handful of systems. In this work, we use a larger sample of randomly-selected halos from the Millennium-II Simulation \citep{bk09} with masses corresponding to $\sim L^*$ galaxies and poor galaxy clusters.

\section{Simulations}\label{simulation}

\begin{figure}[tbp]
\centering % \begin{center}/\end{center} takes some additional vertical space
\includegraphics[width=.49\textwidth]{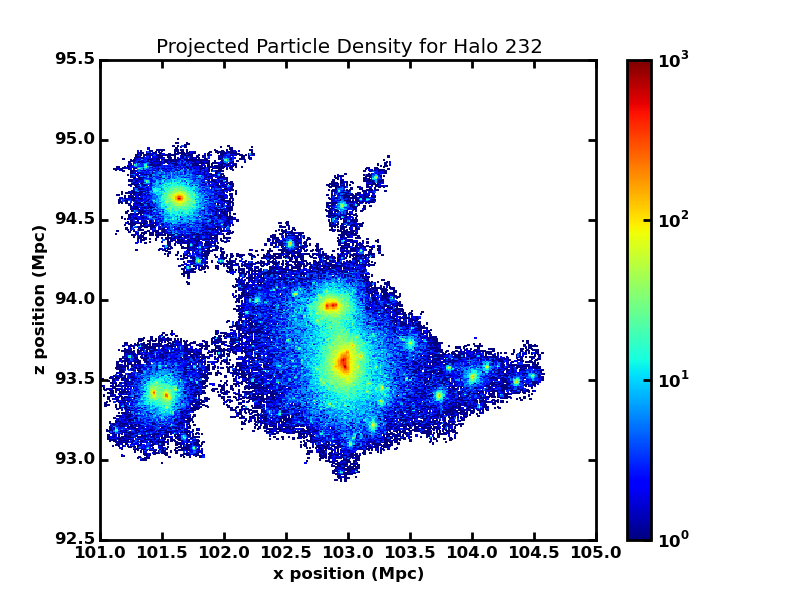}
\hfill
\includegraphics[width=.49\textwidth]{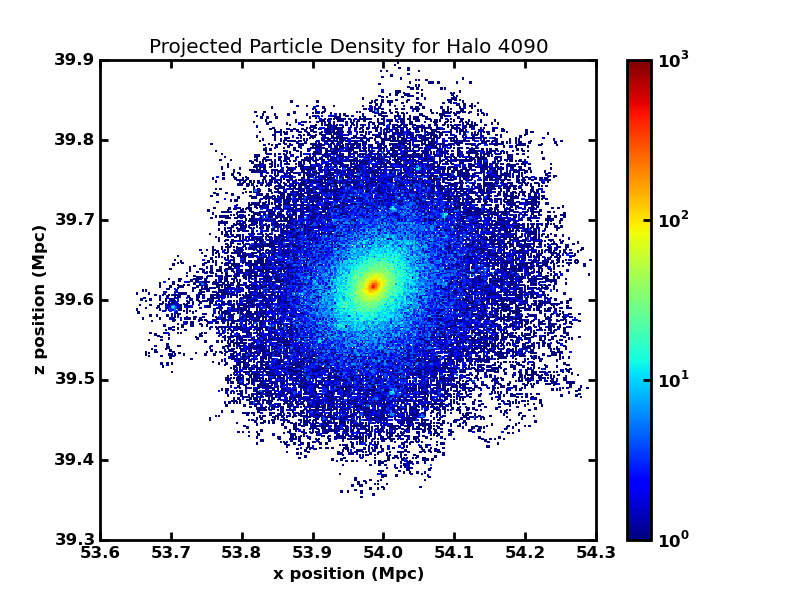}
\caption{ An $x$-$z$ projection of particle positions for halo 232 (left) and 4090 (right), as examples of an out of equilibrium, and equilibrium halos, respectively. In halo 232 multiple structures can be seen merging or passing through one another.  The color represents log of projected density of particles.}
\label{fig:positions}
\end{figure}

Our analysis is based on data from the Millennium-II Simulation \citep{bk09}. The whole simulation is a cube of about $100 h^{-1}\,$Mpc on the side with a spatial resolution (Plummer-equivalent force softening) of $1 h^{-1}\,$kpc. The background cosmology is the concordance $\Lambda$CDM model, with $\Omega_m=0.25$, $\Omega_\Lambda=0.75$, $h=0.73$, and $\sigma_8=0.9$.  The simulation follows the collisionless evolution of $2160^3 \approx 10^{10}$ particles, each with a mass of $6.885\times10^6 h^{-1} M_\odot$.  Gravitationally bound systems are identified by a friends-of-friends algorithm with a linking length of 0.2 times the mean inter particle separation. While many halos are relatively isolated, some are not. Some `halos', especially at the high mass end, consist of multiple merging systems. The merging halos have regions of large densities of particles far from the deepest potential, often in the form of large substructures or a completely separate structure that is in the process of merging or passing near or through the main structure. Figure~\ref{fig:positions} shows all the particles in one such halo file. In these cases we took the dominant mass clump as our halo. For more details on the Millennium-II Simulation and its data products, see \cite{bk09}.

To test DARKexp, we want to fit halos in a wide range of masses. In practice, very high mass halos, $\gtrsim\!10^{14}M_\odot$, are rare, while very low mass systems, $\lesssim\!10^{11}M_\odot$, have too few particles for a meaningful fit. Our data set includes two 100-halo sets with disjointed mass ranges, all at $z = 0$, with total masses ranging from $7.6\times10^{11}M_\odot$ at the low mass end to $4.5\times10^{13}M_\odot$ at the high mass end. The 100 high mass systems were sampled with a random subset of $1/10^{th}$ of the particles in order to keep the file size manageable. The particle masses for these halos were assumed to be $10\times$ larger to compensate in density and potential calculations.

\section{Determining Equilibrium}\label{deteq}

The virial radius of each halo, $r_{200}$, is calculated as the radius within which the average density is $200\rho_{crit}$, and $\rho_{crit}=3H_0^2/(8\pi G)$, where $H_0$ is present day Hubble constant. The halo center is defined as the location of the particle with the deepest potential, $\vec r_{\rm{deep}}$. Virial radii for our low mass halos range from 170~kpc to 280~kpc, and from 370~kpc to 580~kpc for the high mass halos. 

Our analysis requires an estimate of how relaxed a halo is, because DARKexp, being a maximum entropy state, applies only to relaxed systems. Separating equilibrium systems has been discussed in a number of papers \citep{net07,lud12}. Three criteria are generally used: separation between the deepest potential and the center of mass has to be small, the virial ratio of the system has to be close to 1, and the substructure fraction has to be low. The actual values for these criteria have been established by \cite{net07}, and we use their values. We use only the first two criteria.

According to \cite{net07} a system is in equilibrium if 
\begin{equation} 
s = \frac{|\vec r_{\rm{deep}} - \vec r_{\rm{CoM}}|}{r_{200}}\leq 0.07,\label{eqn:sparam}
\end{equation}
where the center of mass of the system, $\vec r_{\rm{CoM}}$ is calculated based on particles within the virial radius.

To calculate the virial ratio, $-2T/P$, we need to calculate total kinetic and potential energies of each system. Obtaining kinetic energy requires us to remove the halo's bulk motion, which we found by averaging the velocities of the central $\sim$10\% of the particles, corresponding to approximately 7\% of the total radial extent of the halo. The resulting bulk velocities are rather insensitive to the exact fraction of particles used. 

\begin{figure}
\centering
\includegraphics[width=.75\textwidth]{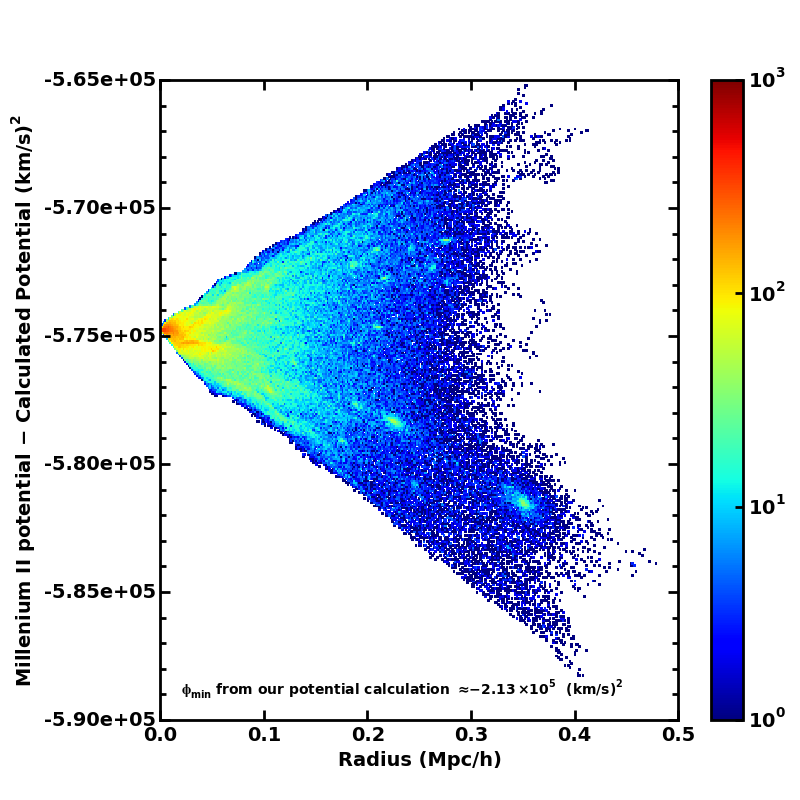}
\caption{The vertical axis shows Millennium-II potential (based on all particles in the simulation) minus the potential calculated by us based on particles within the halo. The scatter at large radii is due to the influence of particles outside of the halo, which were included (not included) in the MII (our) potential calculations. The color shows the log of the particle density.}
\label{fig:potdiff}
\end{figure}

The potential energies of the particles supplied by the Millennium-II simulation team are calculated based on all the particles in the simulation, including those far from the halo in question. How much do these distant particles contribute to the potential energies of particles in a given halo?  In an isolated halo, assumed by DARKexp derivation, or a halo embedded in a uniform infinite medium this contribution is either zero or a constant value. To estimate this contribution in Millennium-II halos, we calculated particle potential energies based on just the particles in the halo, using the smoothing kernel defined in \cite{gadget}. Figure~\ref{fig:potdiff} shows the difference between our calculated potentials and the corresponding Millennium-II values for one halo. For the particles at the deepest potential the difference is just an offset (which we discuss in the next paragraph). At larger radii, there is scatter due to the effects of more distant particles not included in our halo data set. However, this scatter did not significantly affect the estimate of the virial ratio, and the main results of the paper, and so from now on we use Millennium-II particle potentials based on the whole simulation. 

\begin{figure}
\centering
\includegraphics[width=.75\textwidth]{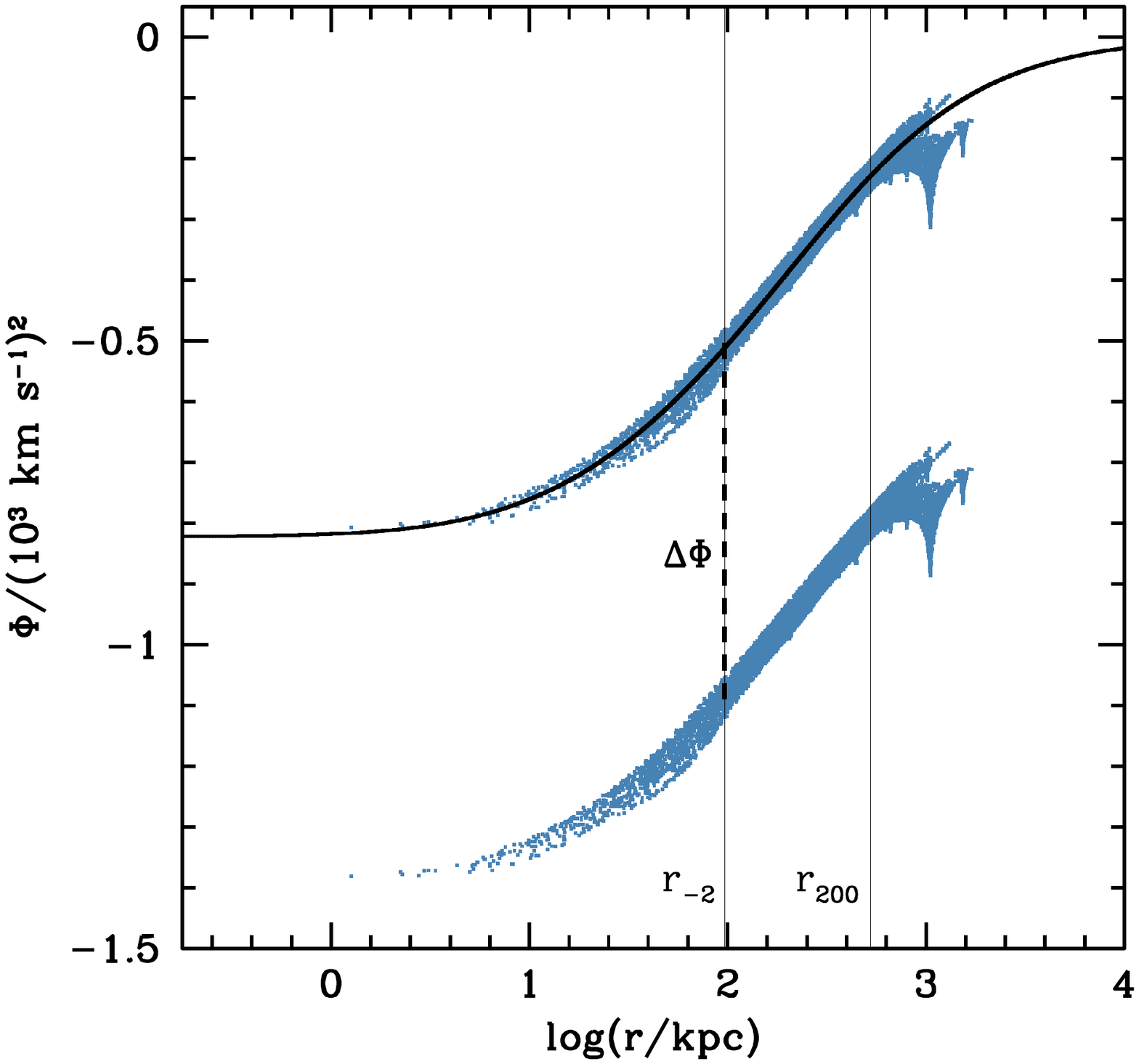}
\vspace{-90pt}
\caption{Particle potential energies as a function of their distance from the halo center, for one example halo. The lower set of blue points are based on all the particles in the Millennium-II simulation. The thick solid curve is the potential of the DARKexp halo fitted to that halo. The potential offset applied to the lower set of points is the difference between $\Phi_{\rm MII}$ of particles found within a few percent $r_{-2}$ and $\Phi_{\rm DARKexp}$ at the same radius. The offset is shown as a thick dashed line, and the shifted particle potentials as the upper set of blue points.}
\label{fig:poten207}
\end{figure}

The systematic offset of the Millennium-II particle potentials needs to be addressed separately. The lower set of light blue points in Figure~\ref{fig:poten207} are the Millennium-II potential values for one example halo, while the thick solid black curve is the potential of DARKexp fitted to the density profile of that halo. The potential offset we applied to the lower set of points is the difference between $\Phi_{\rm MII}$ of particles found within a few percent of $r_{-2}$, and $\Phi_{\rm DARKexp}$ at the same radius. The offset is shown as a thick dashed black line, and the shifted particle potentials as the upper set of light blue points.

Following \cite{net07} we discard systems with $-2T/P>1.35$; we also eliminate a small handful of systems with $-2T/P<0.9$. Between the two criteria, equation~\ref{eqn:sparam} and the virial ratio, 56 of the original 200 halos are eliminated, leaving us with a sample of 144 equilibrium systems.

\section{Fitting Density Distributions}\label{fittingD}

\subsection{Individual Halos}

To be uniform, the fitting has to be done using the same radial range in all halos, defined in terms of a halo's virial radius. We obtained the density distributions of Millennium-II halos by binning particles in logarithmic, radial, spherically symmetric bins, centered on the halo center. Because of softening, the inner radius to be fit is set to $r_{in}=1h^{-1}$kpc. The outer radius is $r_{out}=0.5r_{200}$. We exclude the outermost $0.5r_{200}$ because many systems have substructures in that region, resulting in spikes in the density profile which bias the fitting.

The fitting of the density profiles should be done in three dimensional parameter space of $\phi_0$, and scalings in $r$ and $\rho(r)$, but because the DARKexp density distribution does not have an analytic form,  we used a discrete set of pre-calculated density profiles with $\Delta \phi_0 = 0.1$ increments. For a given $\phi_0$, the density profile fitting was done using a Markov Chain Monte Carlo method with Gaussian random steps, in log-log space of two parameters: offsets in $\log(r)$ and $\log(\rho)$. For each step and each set of fitting parameters, the root-mean-square deviation was calculated using:
\begin{equation}
RMS=\sqrt{{\sum_i\Bigl\vert \log \left[\rho_{\rm {MII}}(i)\right] - \log \left[\rho_{\,\rm {DARKexp}}(i)\right] \Bigr\vert^2}},
\label{eq:DensityRMSformula}
\end{equation}
where the sum is over 20-25 radial bins inside the range $r_{in}$ to $r_{out}$. To avoid wasting time on parameters that would not produce a good fit, $\phi_0$ was limited to a range between $0.5$ and $8.0$. Systems with very small $\phi_0$ have large flat density cores, inconsistent with simulated halos, while those with $\phi_0\sim 8$ or larger have slopes interior to $r_{-2}$ (radius where $d\log\rho/d\log r=-2$) of about $-2$ \citep{hjorth15}. Steeper central density profiles are not expected in simulated halos. The resulting $\phi_0$ values are called $\phi_{0,D}$ to denote that they come from density fitting.

To check the robustness of our density fitting we perform an independent fitting to the density profile slopes, $\gamma(r)=-d\log[\rho(r)]/d\log(r)$. These yielded very similar results, with only four of 144 equilibrium halos (see Section~\ref{deteq}) giving $|\phi_{0,D}-\phi_{0,D,{\rm slope}}|>0.8$. Excluding the 4 outliers, the dispersion, $\Bigl[\langle(\phi_{0,D}-\phi_{0,D,{\rm slope}})^2\rangle\Bigr]^{1/2}=0.28$, is taken as the measure of the uncertainty in fitting the density profile shape. 

\begin{figure}
\includegraphics[width=.435\textwidth]{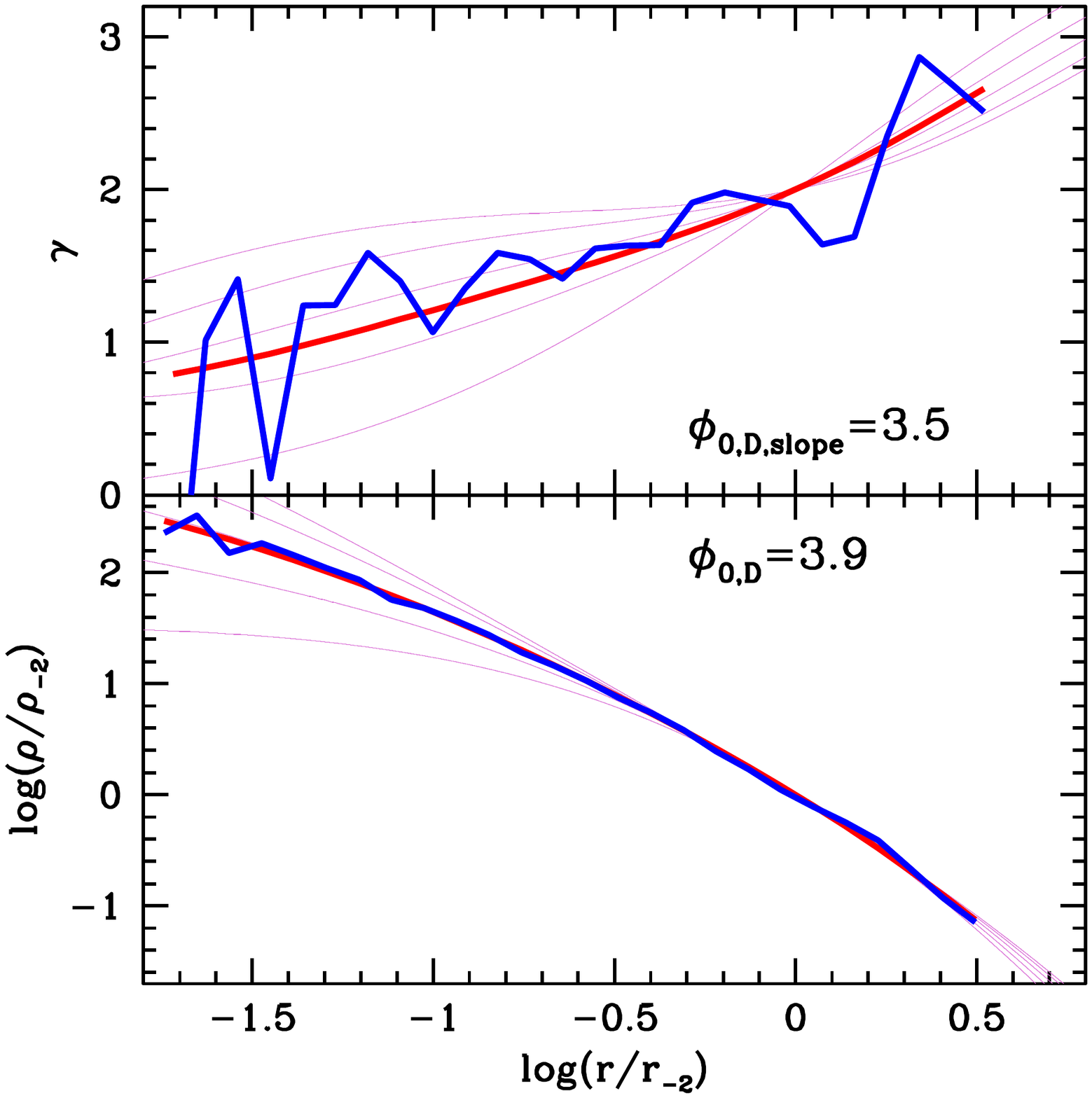}
\vspace{-75pt}
\includegraphics[width=.435\textwidth]{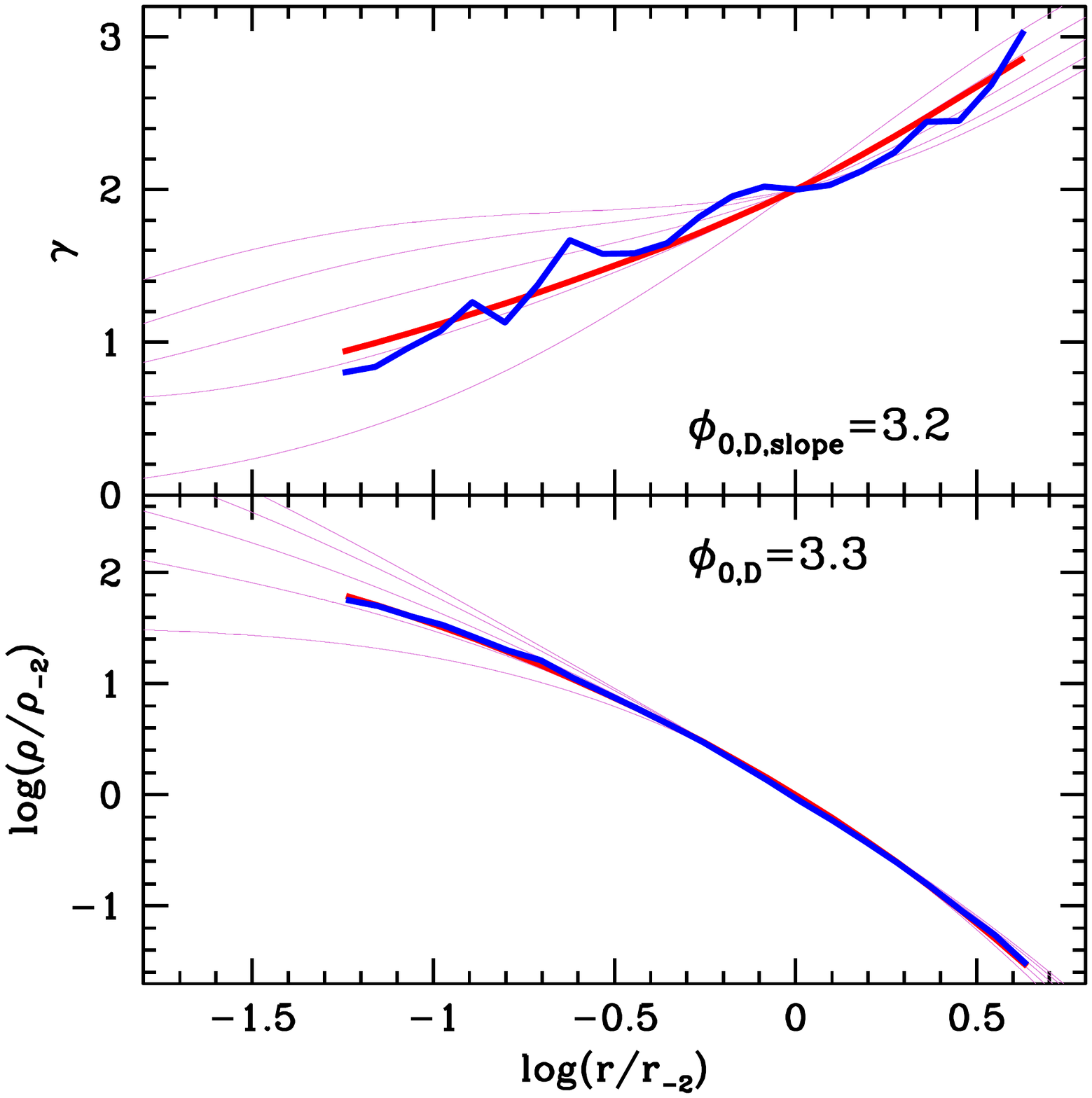}
\vspace{-75pt}
\includegraphics[width=.435\textwidth]{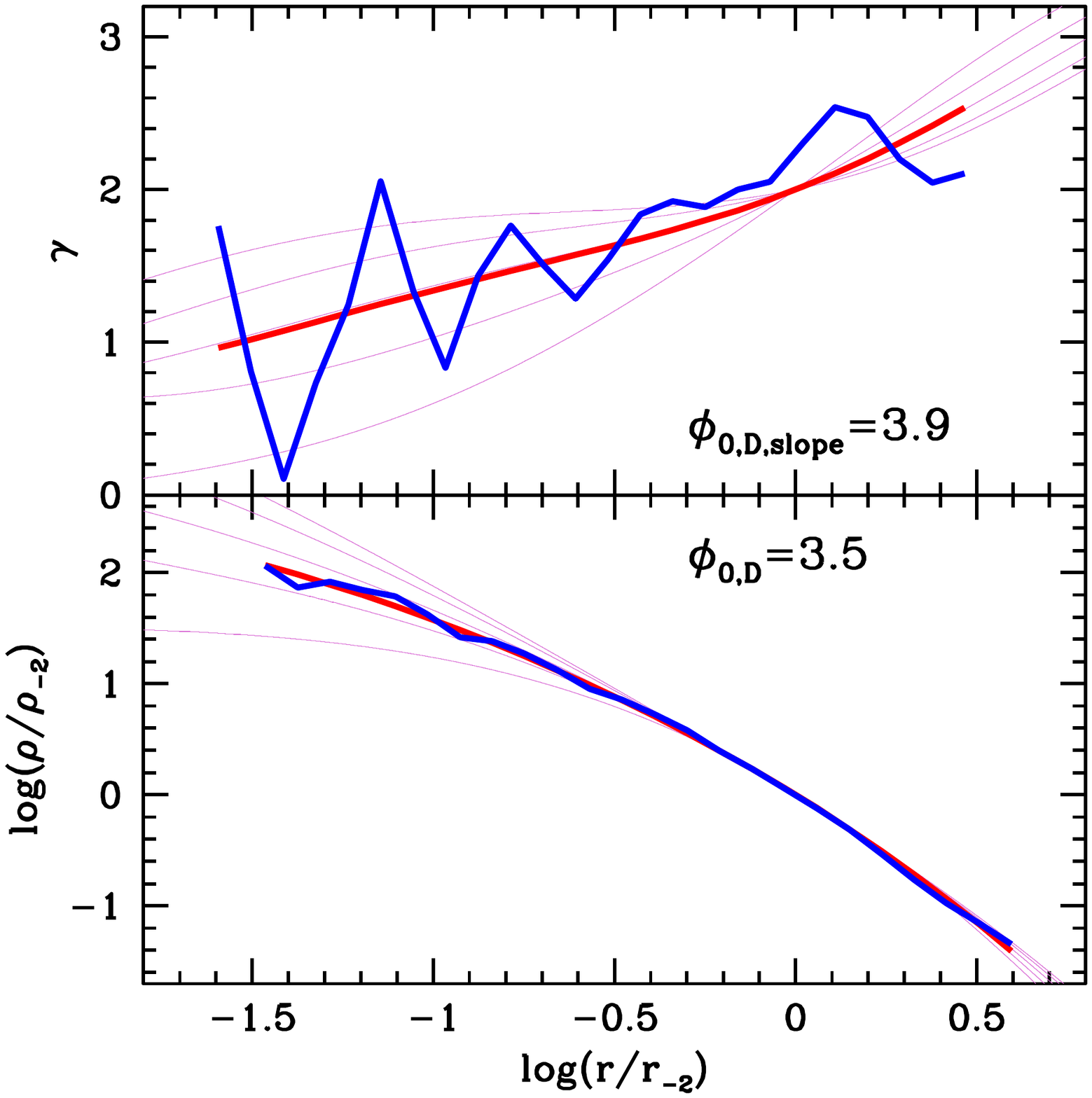}
\hfill
\includegraphics[width=.435\textwidth]{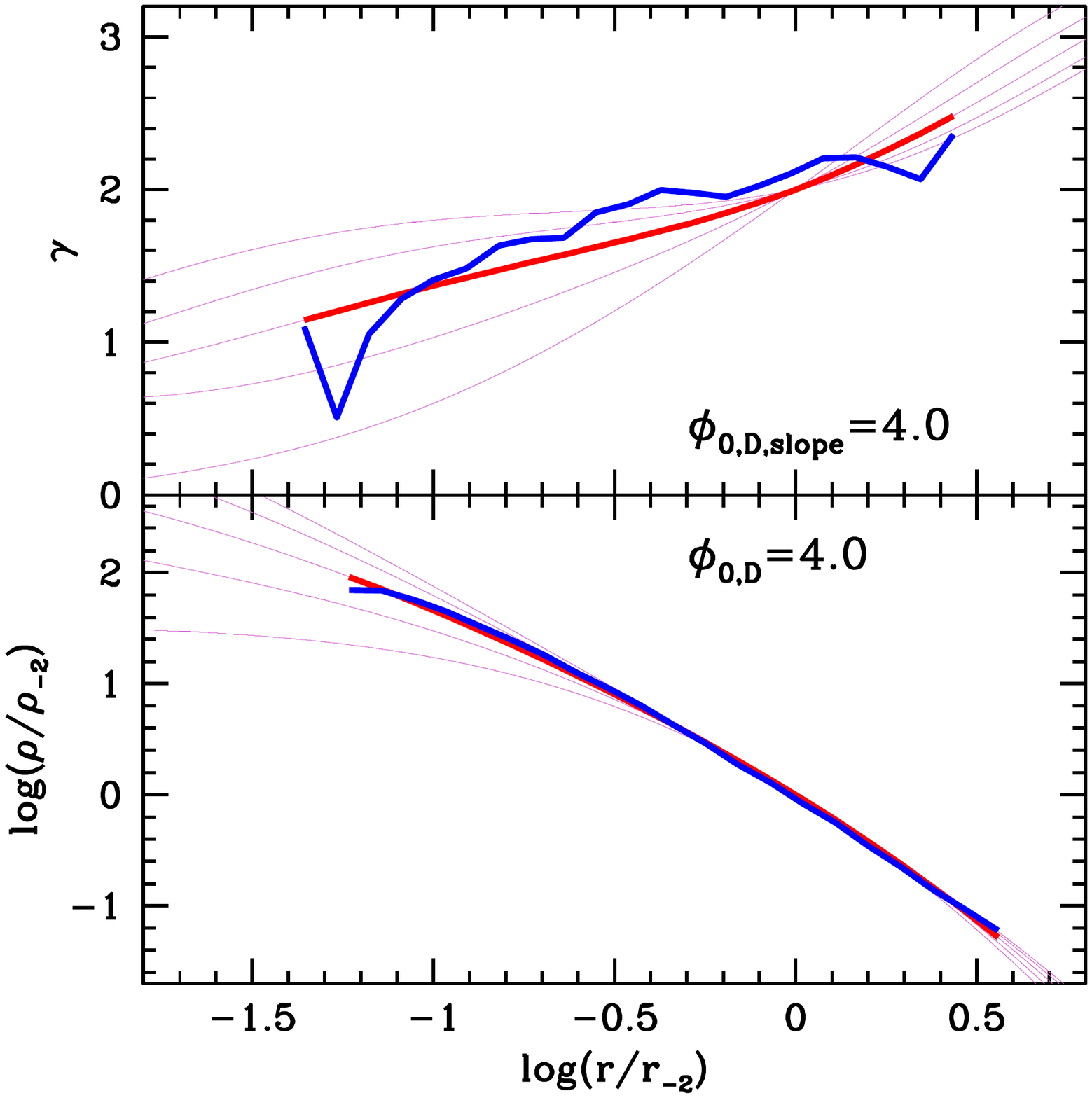}
\vspace{-75pt}
\includegraphics[width=.435\textwidth]{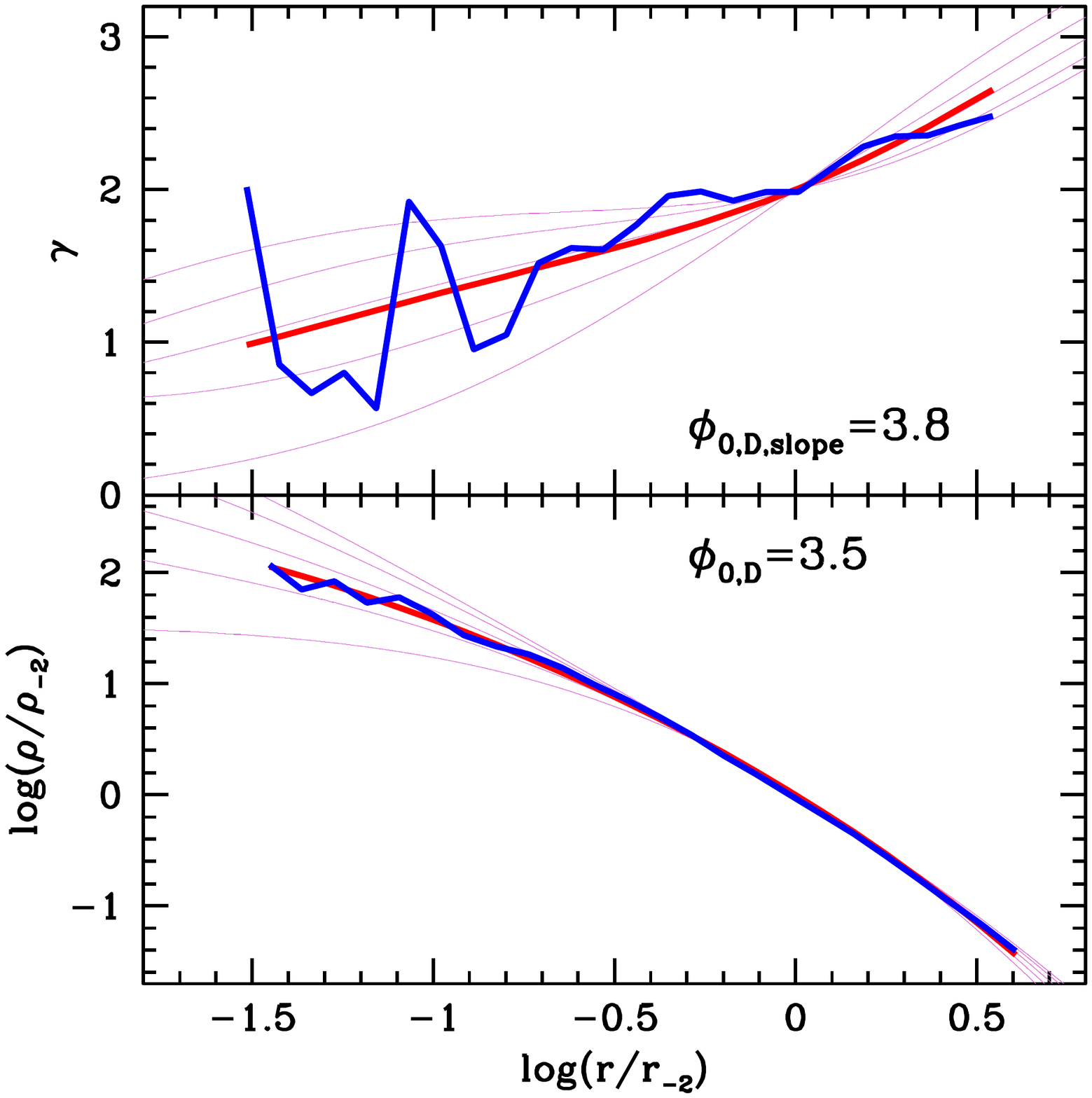}
\hfill
\includegraphics[width=.435\textwidth]{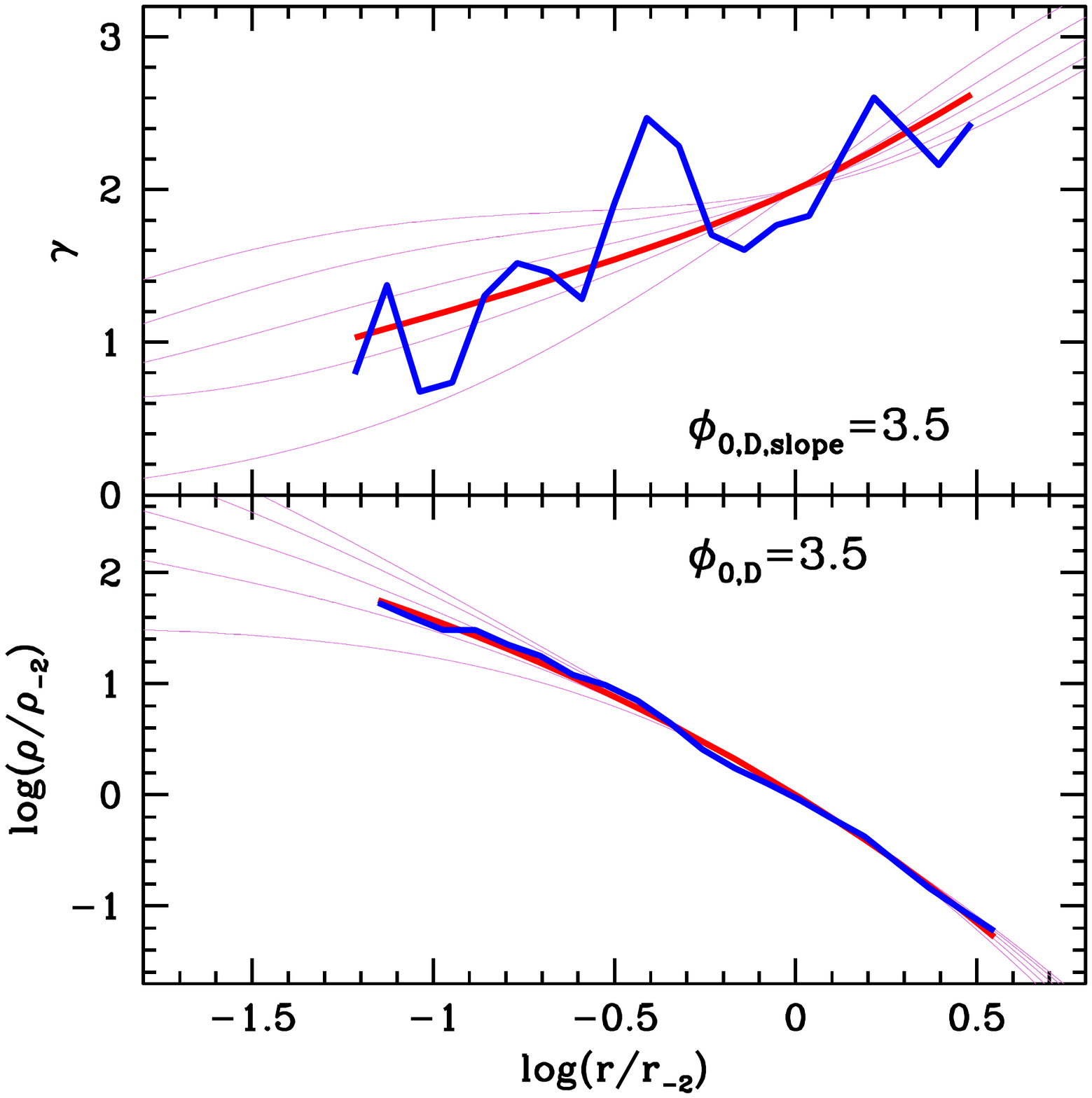}
\vspace{+18pt}
\caption{Fits to log of density, $\log[\rho(r)]$, and density slope, $\gamma(r)$, profiles for selected high mass (left panels) and low mass (right panels) halos, presented in the same order as in Fig.~\ref{fig:energyfits}. The halos were chosen based on the quality of their $N(E)$ fits (Section~\ref{indivhalos}). The thick smooth (red) curves are the best fit DARKexp; Millennium-II data are the (blue) jagged lines. The thin (magenta) curves are DARKexp of $\phi_0=6,5,4,3,2$ (higher $\phi_0$ give rise to steeper central density slopes).}
\label{fig:densityfits}
\end{figure}

Figure~\ref{fig:densityfits} shows examples of six halos, three high mass (left panels) and three low mass (right panels). The upper and lower panels in each of the six sets shows the fits to the slope profile, $\gamma(r)$, and to the log of the density profile, $\log[\rho(r)]$, respectively. The blue jagged lines are the Millennium-II data, while the thick smooth red lines are the DARKexp fit profiles. For reference, the magenta thin curves in each panel represent DARKexp of $\phi_0=6,5,4,3,2$, from top to bottom at small radii, in both $\gamma(r)$ and $\rho(r)$ panels. Unless otherwise stated, we use $\phi_{0,D}$ fitted to the density profile in the analysis.  DARKexp provides very good fits to most Millennium-II halos.

\subsection{Average Halos}\label{avehaloD}

\begin{figure}
\includegraphics[width=.5\textwidth]{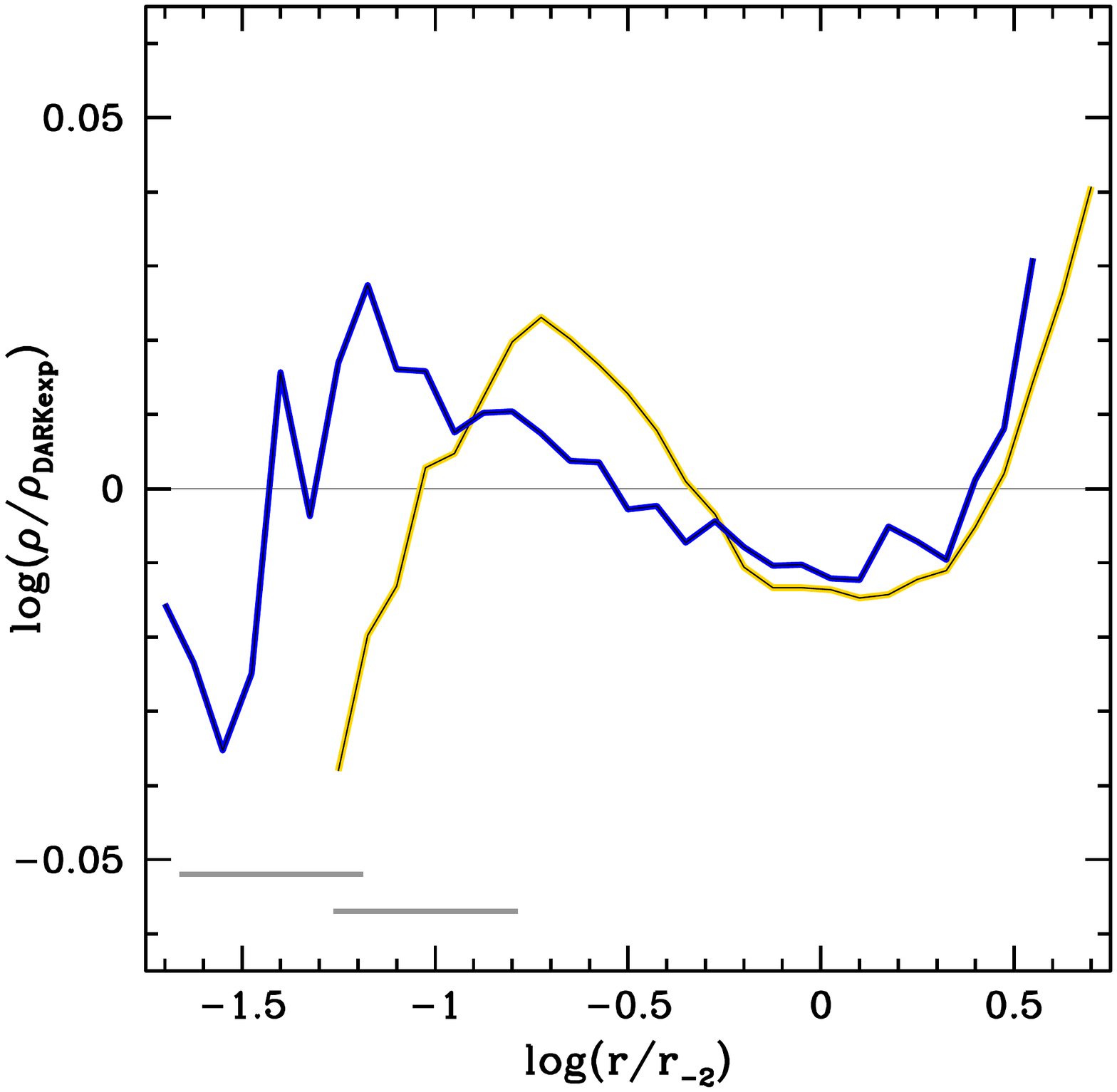}
\vspace{-60pt}
\includegraphics[width=.5\textwidth]{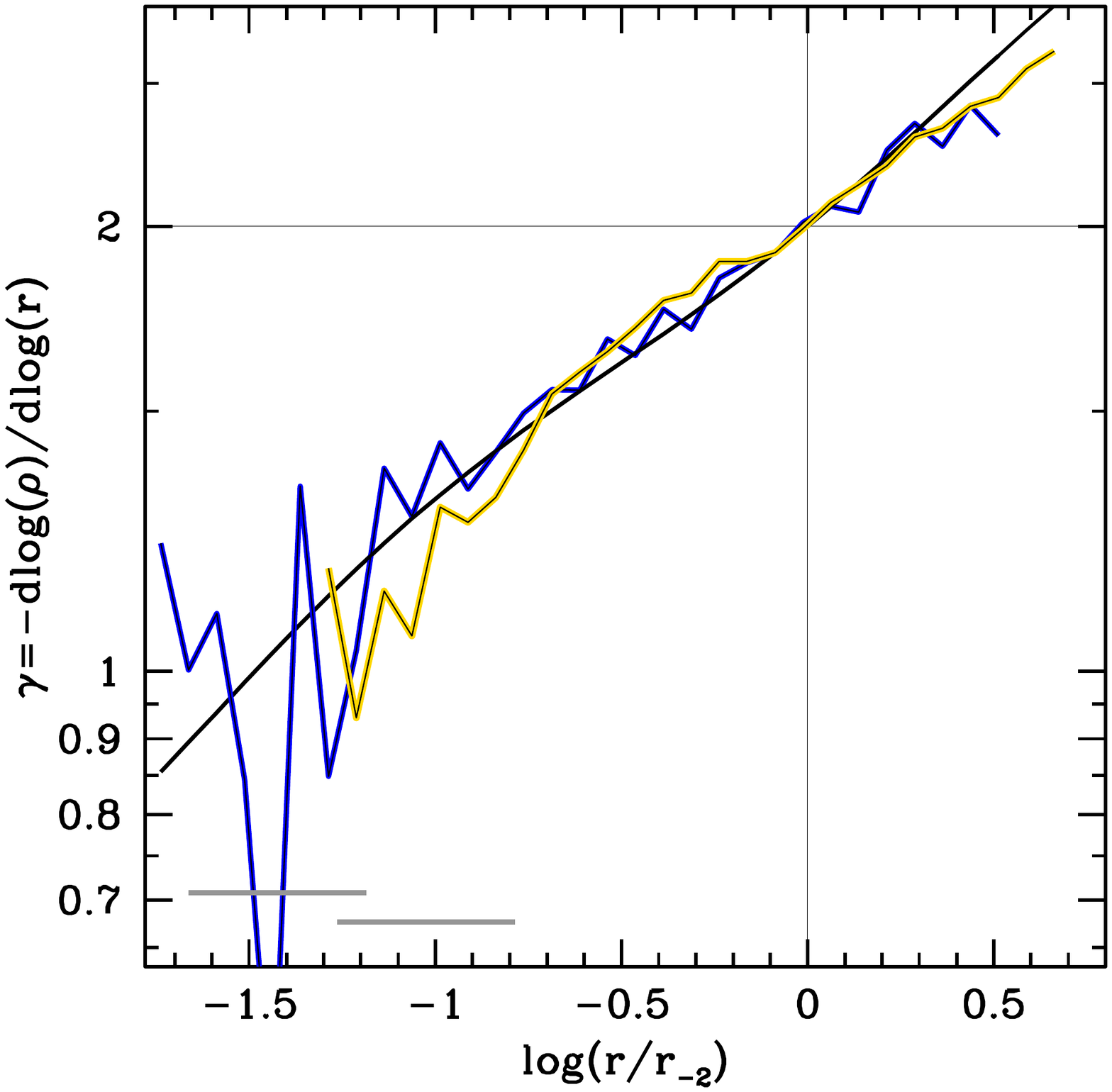}
\caption{{\it Left panel: }The dark blue and the light yellow lines show deviations of the log of the density profile, $\log\rho(r)$, from best-fitting DARKexp for high and low mass halos, respectively. All halos in each of the two groups were averaged to produce these deviation profiles. Two solid gray bars in the lower left of the plot show typical $1\rightarrow 3$ smoothing lengths intervals for the high mass (upper left bar) and lower mass (lower right bar) halos. {\it Right panel: } Similar to the left panel, but here we plot density profile slope as the vertical axis. The black smooth line is DARKexp of $\phi_0=3.8$.}
\label{fig:profdev}
\end{figure}

To reduce noise in the density profiles and make the systematic trends stand out, we divide the halos into two groups, low mass (those with $r_{200}<280$~kpc), and high mass (those with $r_{200}>370$~kpc), and average halos within each group. Direct averaging, i.e. adding up $\log(\rho)$ or $\rho$ from many halos in each radial bin and then dividing by the number of halos, is not the best strategy because each halo is fit by DARKexp with different $\phi_{0,D}$. Instead, we calculate the deviation of each halo's density from its best fit, $[\log\rho_{\rm MII}(r)-\log\rho_{\rm DARKexp,\phi_{0,D}}(r)]$ for each radial bin and average these deviations within each bin. The dark blue and light yellow lines in Figure~\ref{fig:profdev} show these deviations for the high mass and low mass groups, respectively. Note that in the left panel, these lines need not go through $\log(\rho/\rho_{\rm DARKexp})=0$ at $\log(r/r_{-2})=0$, but $\log(\rho/\rho_{\rm DARKexp})=0$ line should roughly bisect the data profile deviations. 

Let us examine the differences between the deviations of the low mass and high mass halos from their best fit DARKexp, shown in the left panel of Figure~\ref{fig:profdev}. The high mass halos (dark blue line) show somewhat smaller systematic deviations compared to the low mass halos (light yellow line). At large radii, $\log(r/r_{-2})\simgt -0.5$, the two agree reasonably well, but at small radii they do not. The peak in the low mass halos is around $\log(r/r_{-2})\approx -0.75$, followed by a drop at smaller radii, while the peak in the high mass halos occurs around $\log(r/r_{-2})\approx -1.2$, also followed by a drop at smaller radii. Assuming typical values for $r_{-2}$ for low and high mass halos, both the peaks occur at about 3 simulation smoothing lengths, or $3\times 1.37=4.11$~kpc. 

At these separations, one expects the density profiles to be flattened by the numerical resolution effects, which is seen in the right panel of Figure~\ref{fig:profdev} as a decrease in slope. The drop mentioned above (left panel of Figure~\ref{fig:profdev}) is this flattening, and the peak is made up of particles that would have found themselves closer to the halo center if the smoothing length were negligibly small. The fact that the peak and the drop are shifted towards smaller radii for the high mass halos, compared to the low mass ones, is consistent with this interpretation, as is the amount of the shift.  If numerical resolution were a negligible effect, one could imagine that the particles in the peak would move to smaller radii, thereby reducing, or eliminating the peak and the drop in the density profile deviations from best-fitting DARKexp. This would also steepen the central profile slope.

\subsection{Velocity dispersion and pseudo phase-space density profiles}\label{sigma}

Because DARKexp assumes velocity isotropy, we concentrate on comparison of quantities that are least sensitive to velocity anisotropy, namely density profiles and differential energy distributions \citep{BT87,paperIII}.

\begin{figure}
\includegraphics[width=.500\textwidth]{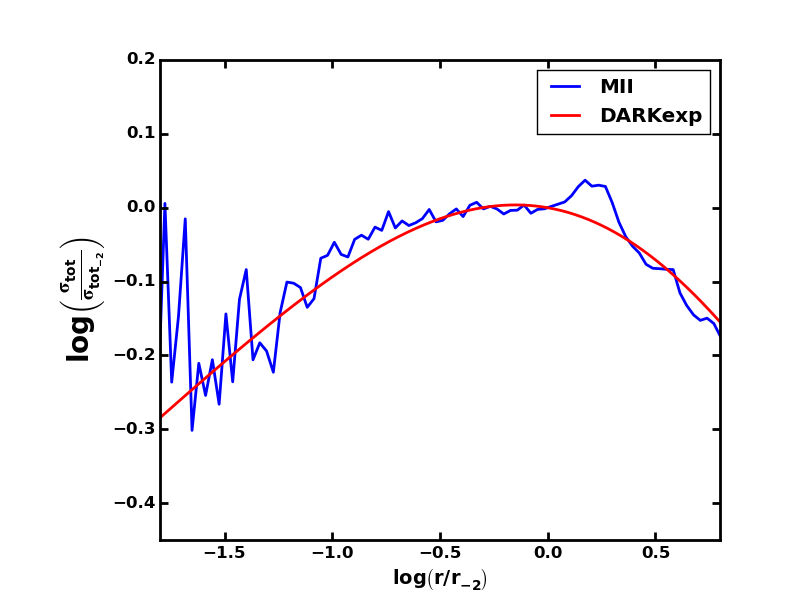}
\vspace{0pt}
\includegraphics[width=.500\textwidth]{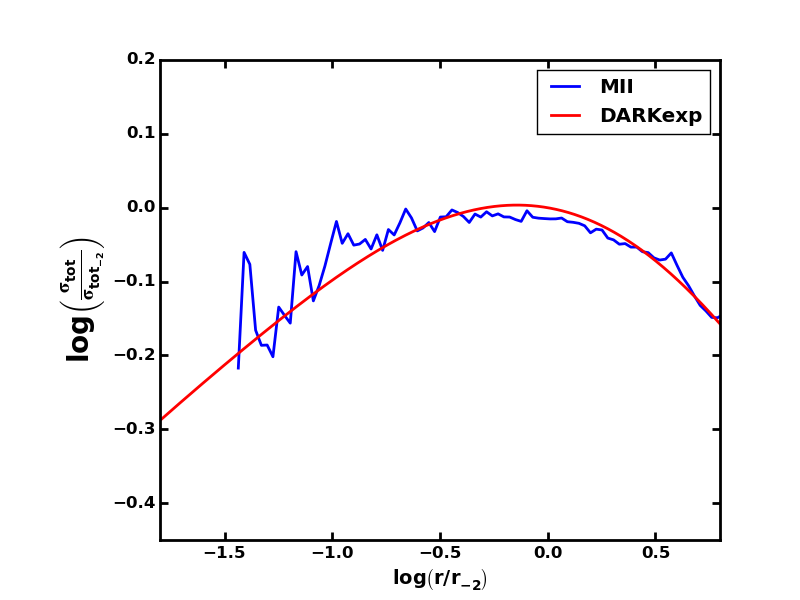}
\vspace{0pt}
\includegraphics[width=.500\textwidth]{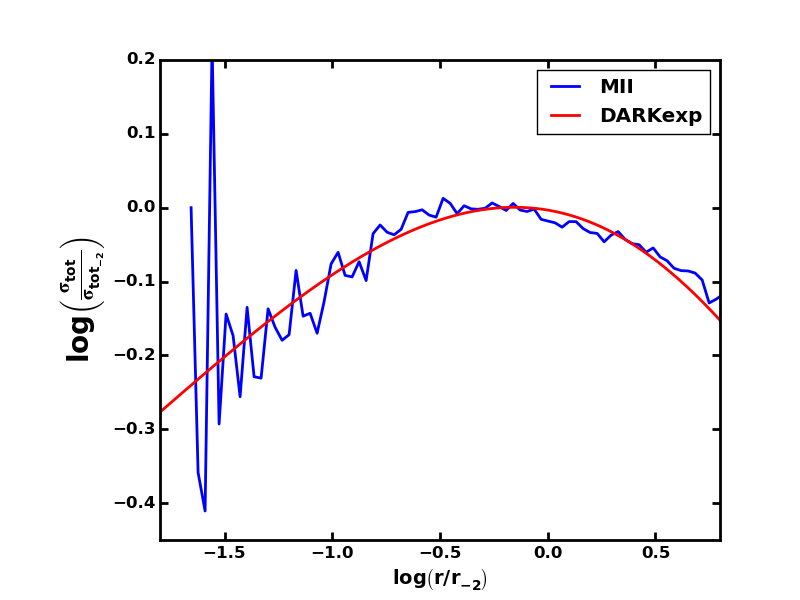}
%\hfill
\includegraphics[width=.500\textwidth]{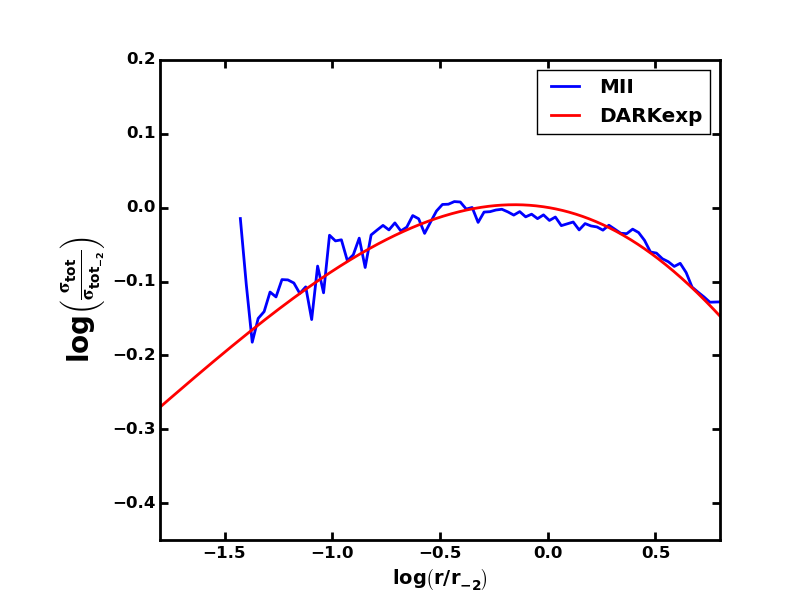}
\vspace{0pt}
\includegraphics[width=.500\textwidth]{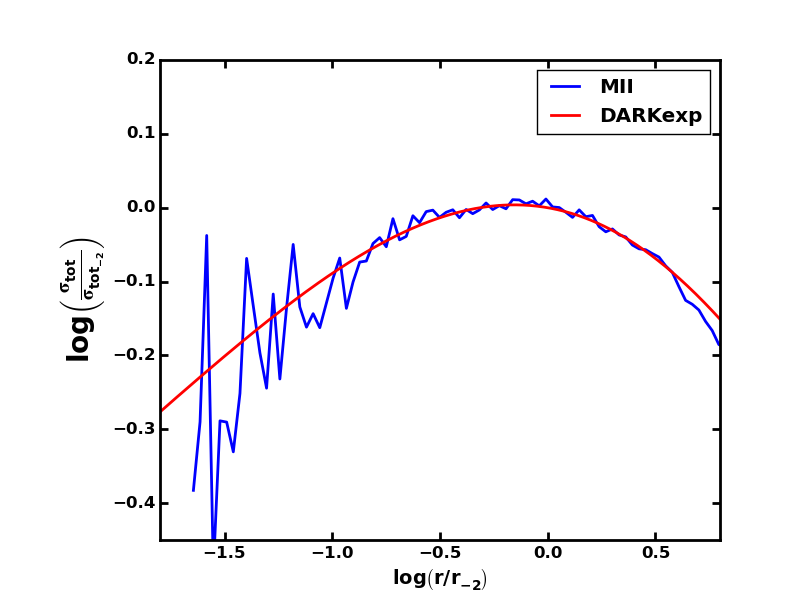}
\hfill
\includegraphics[width=.500\textwidth]{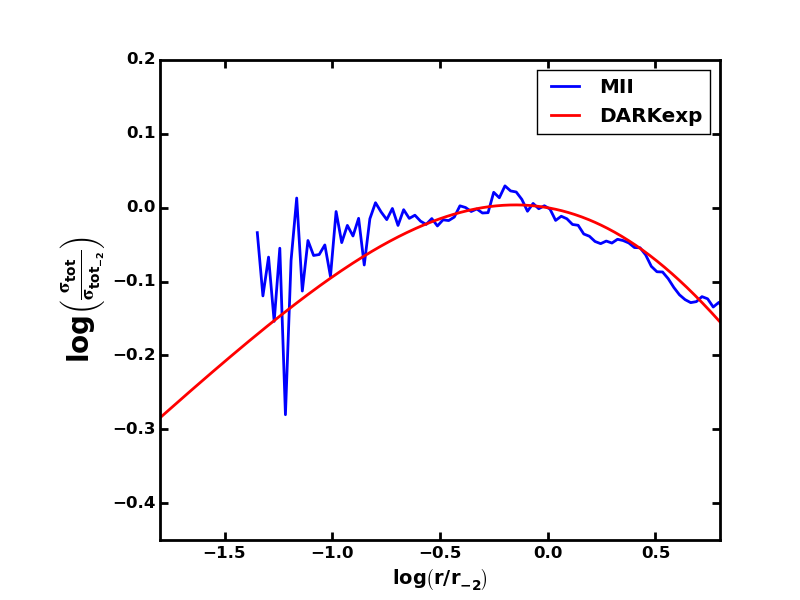}
\vspace{0pt}
\caption{Fits to total velocity dispersion profiles. Both axes as in units of values at $r_{-2}$. The halos shown, and the extent of the horizontal axis are the same as in Figure~\ref{fig:densityfits}.}
\label{fig:sigmafits}
\end{figure}

\begin{figure}
\includegraphics[width=.500\textwidth]{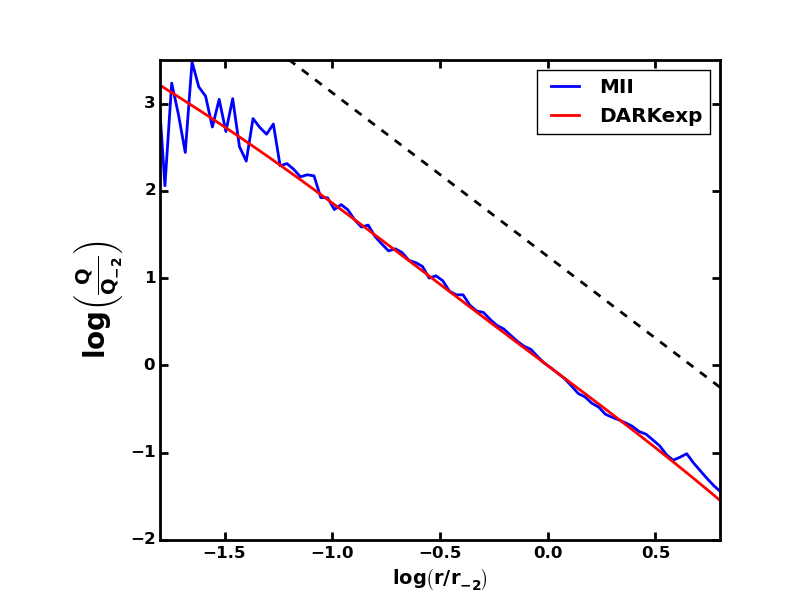}
\vspace{0pt}
\includegraphics[width=.500\textwidth]{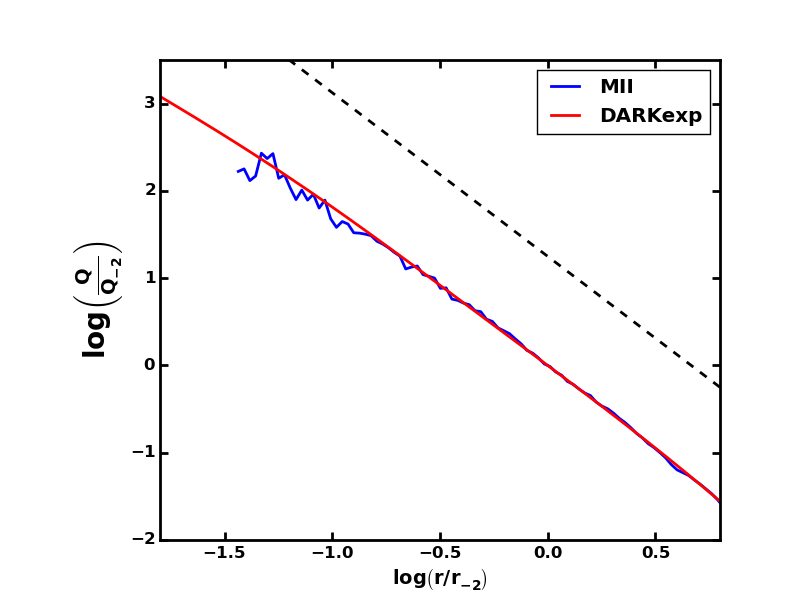}
\vspace{0pt}
\includegraphics[width=.500\textwidth]{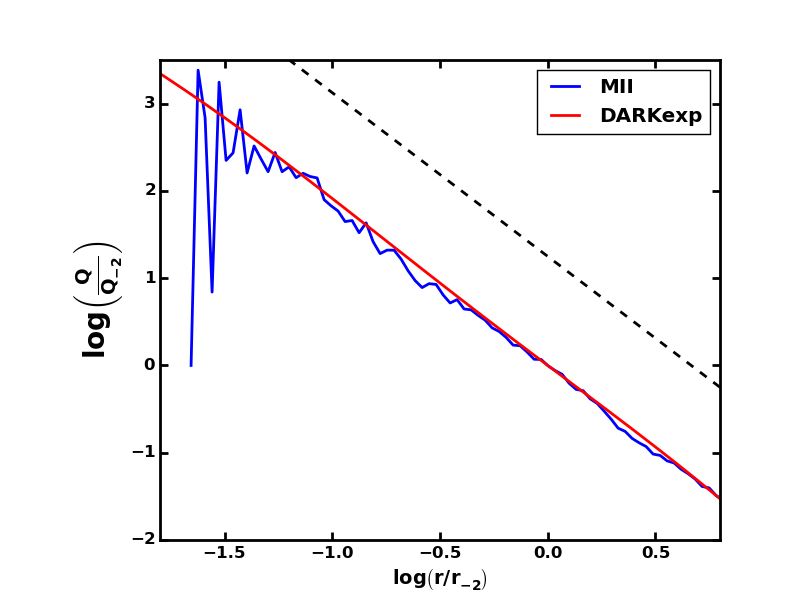}
%\hfill
\includegraphics[width=.500\textwidth]{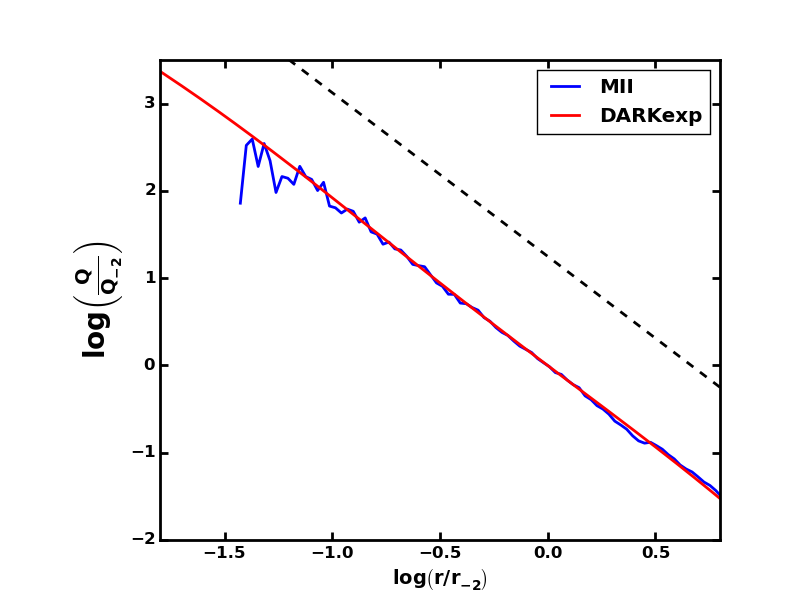}
\vspace{0pt}
\includegraphics[width=.500\textwidth]{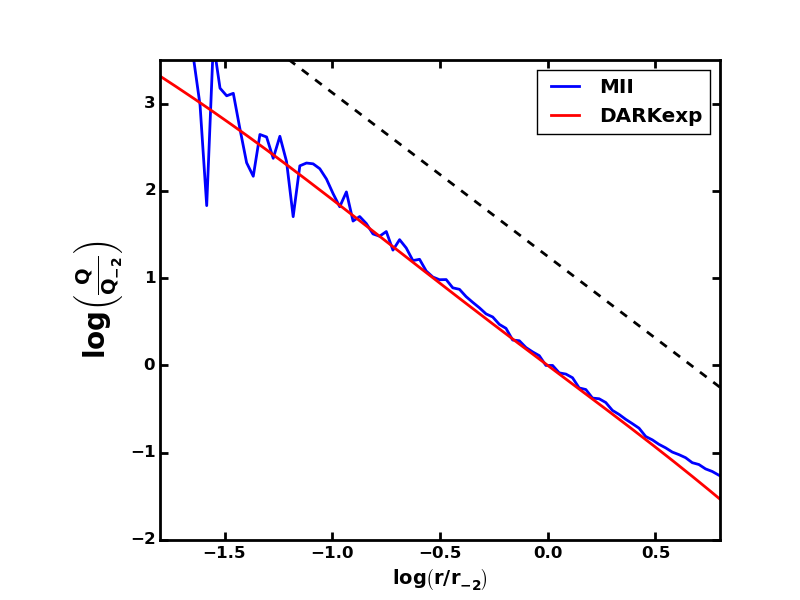}
\hfill
\includegraphics[width=.500\textwidth]{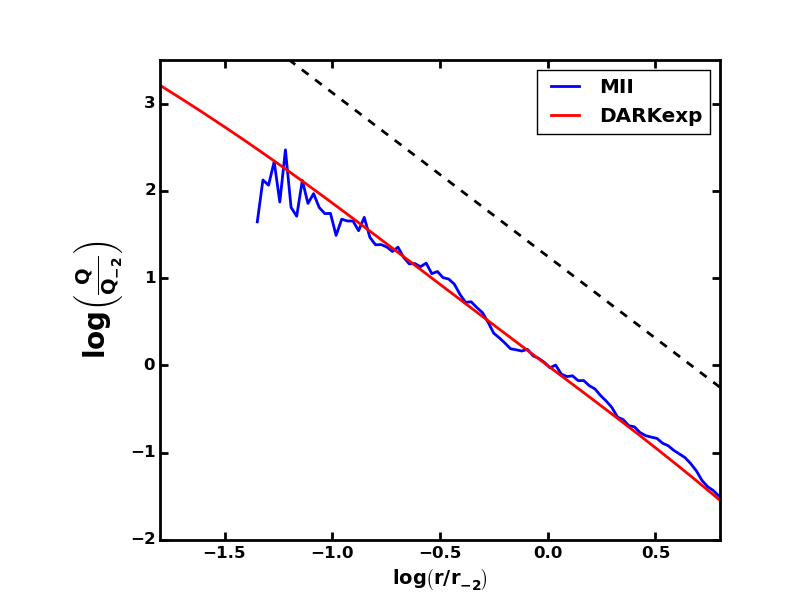}
\vspace{0pt}
\caption{Fits to the pseudo phase-space density profiles, $Q=\rho/\sigma^3$, of the same halos as shown in Figures~\ref{fig:densityfits} and \ref{fig:sigmafits}. Both axes as in units of values at $r_{-2}$. The straight dashed line has a slope of -$d\log(Q)/d\log(r)=-1.875$, and is shown for reference.}
\label{fig:phafits}
\end{figure}

However, for completeness, we also present spherically averaged radial distributions of the total velocity dispersion, $\sigma$, and pseudo phase-space density, $Q=\rho/\sigma^3$, and compare them to the DARKexp predictions, in Figures~\ref{fig:sigmafits} and \ref{fig:phafits}, respectively. Within noise, both agree well with DARKexp. Pseudo phase-space density profiles are observed to be nearly power-laws in simulated and semi-analytical halos \citep{tn01,austin05}, although the reason for this behaviour is not yet known. Figure~\ref{fig:phafits} shows that both Millennium-II halos and DARKexp $Q$ profiles are nearly power laws, with the slope consistent with that of other simulated and semi-analytical halos, $d\log(Q)/d\log(r)=-1.875$, shown as the dashed line in each panel.

\section{Fitting Energy Distributions}\label{fittingE}

\subsection{Individual Halos}\label{indivhalos}

The particle kinetic and potential energies obtained in the previous section are now used to construct the differential energy distribution, $N(E)$, i.e. the number of particles per linear energy bin. Note that while the identification of the correct center of the halo is important for the density profile, it is much less critical for $N(E)$. 

To do the fits, one needs to specify the energy interval where the fit is to be carried out. At the most bound energy end, the distribution of particles truncates very sharply, providing a natural lower limit on energy, $E_{\rm lo}$. At the high energy end, we choose the energy corresponding to the potential energy at $r_{200}$, i.e. $E_{\rm hi}=\Phi(r_{200})$. This choice, though well motivated, is still somewhat arbitrary, and will miss some of the halo particles that are bound to the halo, but whose total energies are larger than $\Phi(r_{200})$. We will return to this in Section~\ref{comp}.

With these energy limits, the energy distribution fitting was done using MCMC in three dimensional parameter space of $A$, $\beta$, and $\Phi_0$ of eq.~\eqref{DARKexpEnergy}. Initial values were chosen so that the initial $\phi_0 = \beta\,\Phi_0 = 3.0$ and the chain was constrained to the same range as in the density fitting: $0.5\leq\phi_0\leq8.0$. The RMS at each step in the chain was calculated using:
\begin{equation}
RMS = \sqrt{{\sum_i \Bigl\vert \log \left[N(i)\right] - \log\left[N_{\rm{DARKexp}}(i)\right] \Bigr\vert^2}},
\label{eq:EnergyRMSformula}
\end{equation}
where the sum is over the energy bins. We chose not to weigh by the particle number in each bin for two reasons. First, the fluctuations in the energy distribution $N(E)$ are not dominated by Poisson noise in the number of particles, but mainly by the substructure, and other density perturbations within the halos. Second, the tightly bound end of the energy distribution, the `$-1$' part of eq.~\eqref{DARKexpEnergy}, is an important aspect of the $N(E)$ shape of the DARKexp model, but which contains relatively few particles. 

\begin{figure}
\includegraphics[width=.42\textwidth]{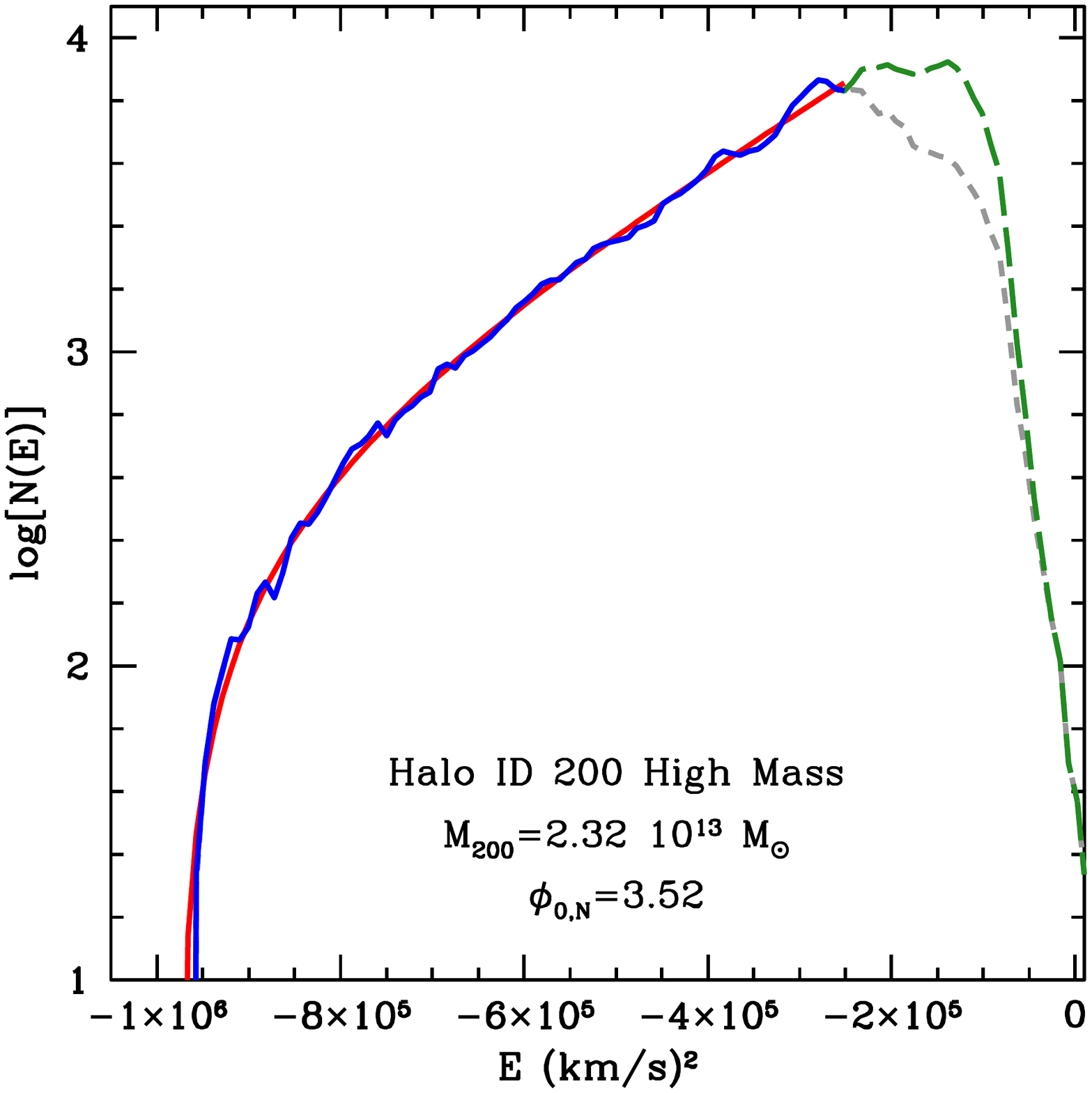}
\vspace{-75pt}
\includegraphics[width=.42\textwidth]{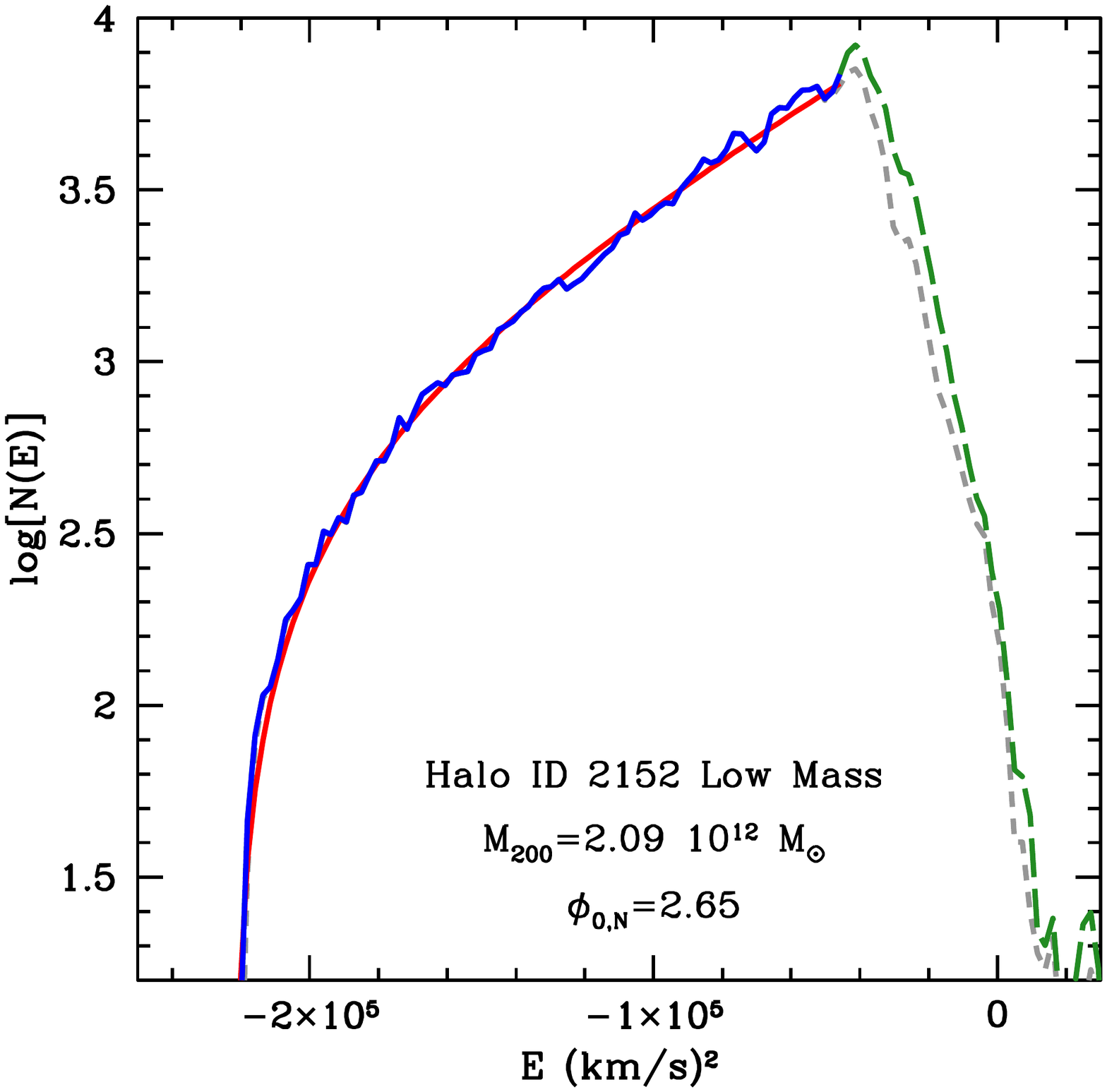}
\vspace{-75pt}
\includegraphics[width=.42\textwidth]{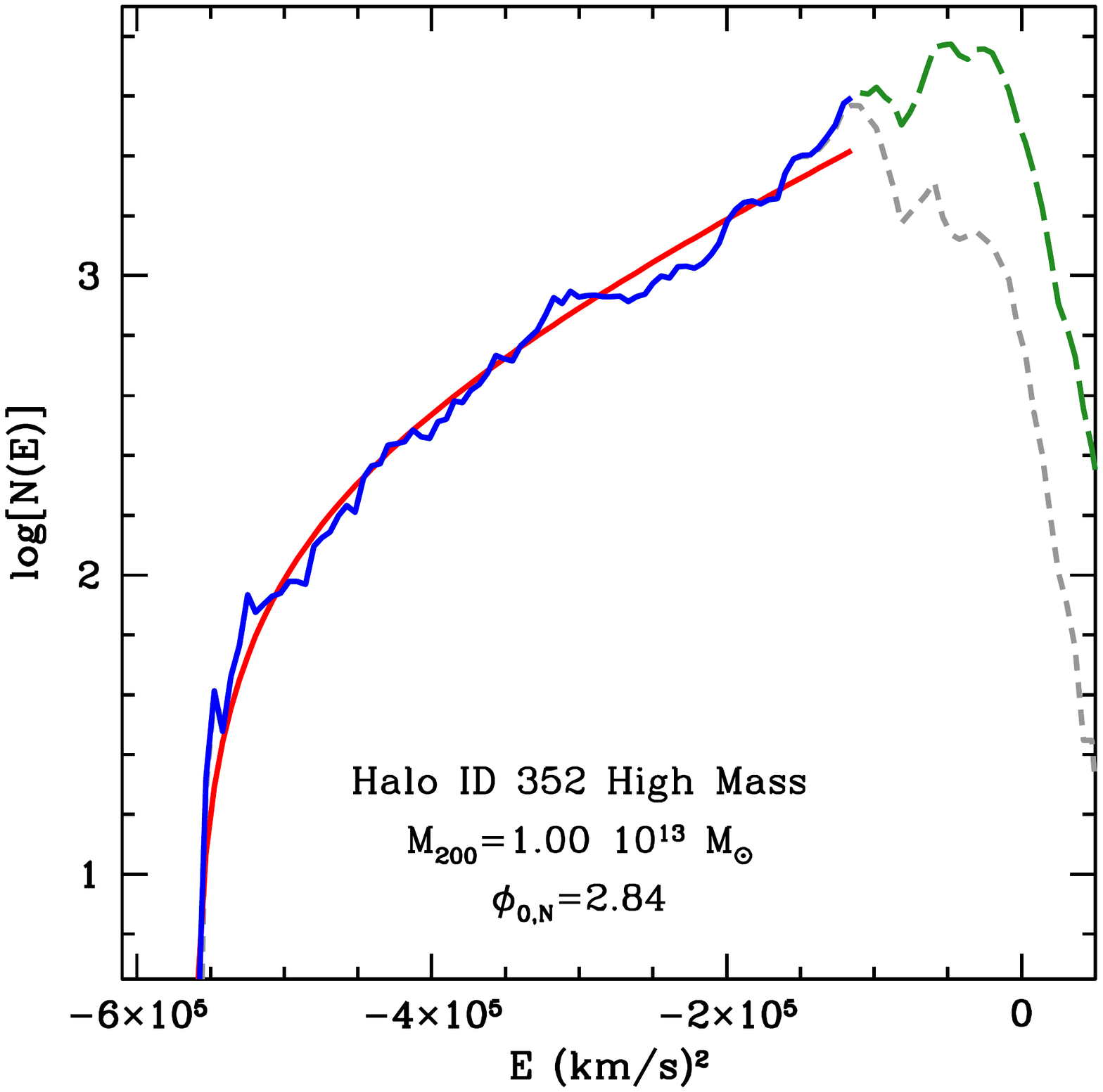}
\hfill
\includegraphics[width=.42\textwidth]{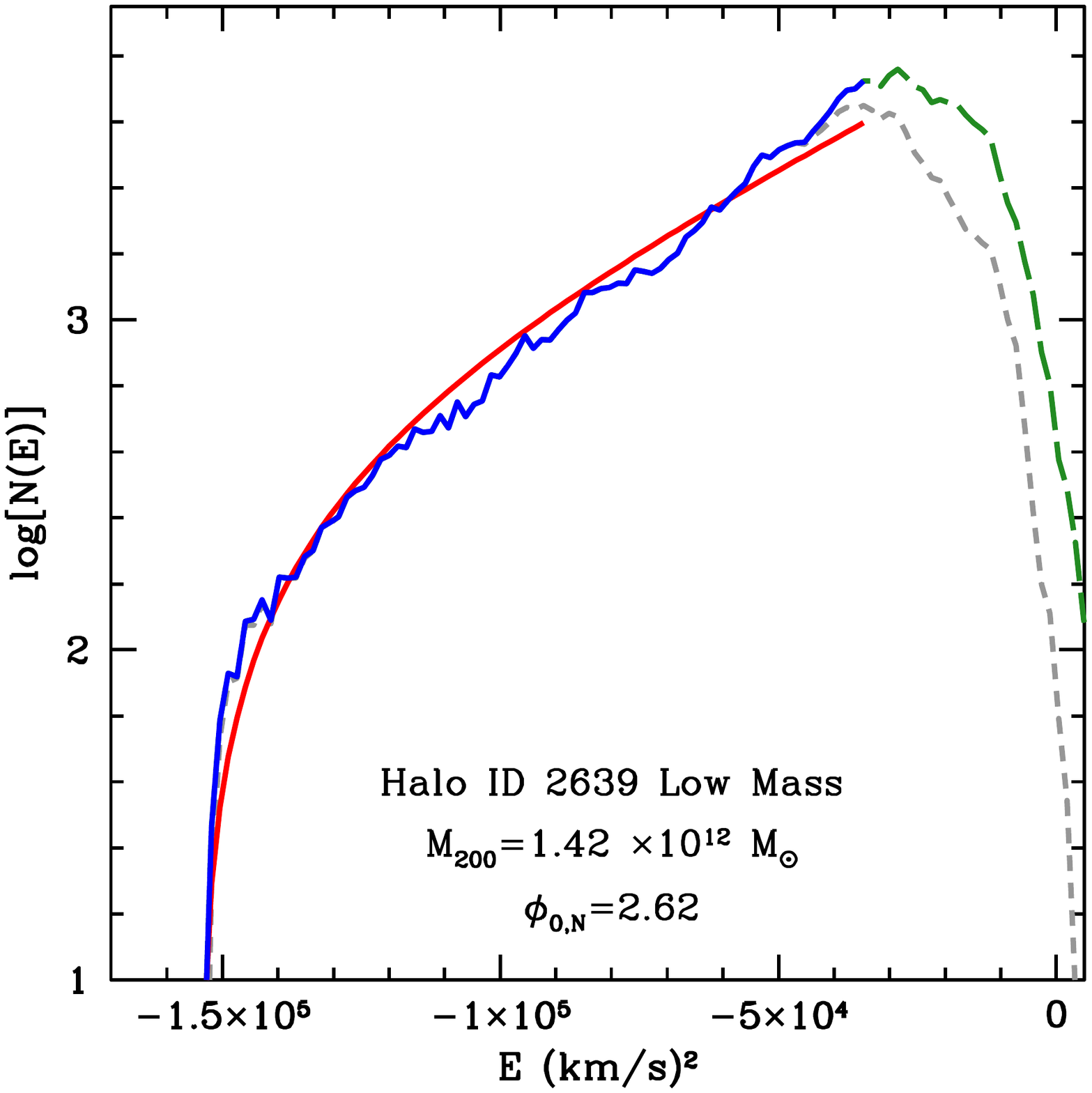}
\vspace{-75pt}
\includegraphics[width=.42\textwidth]{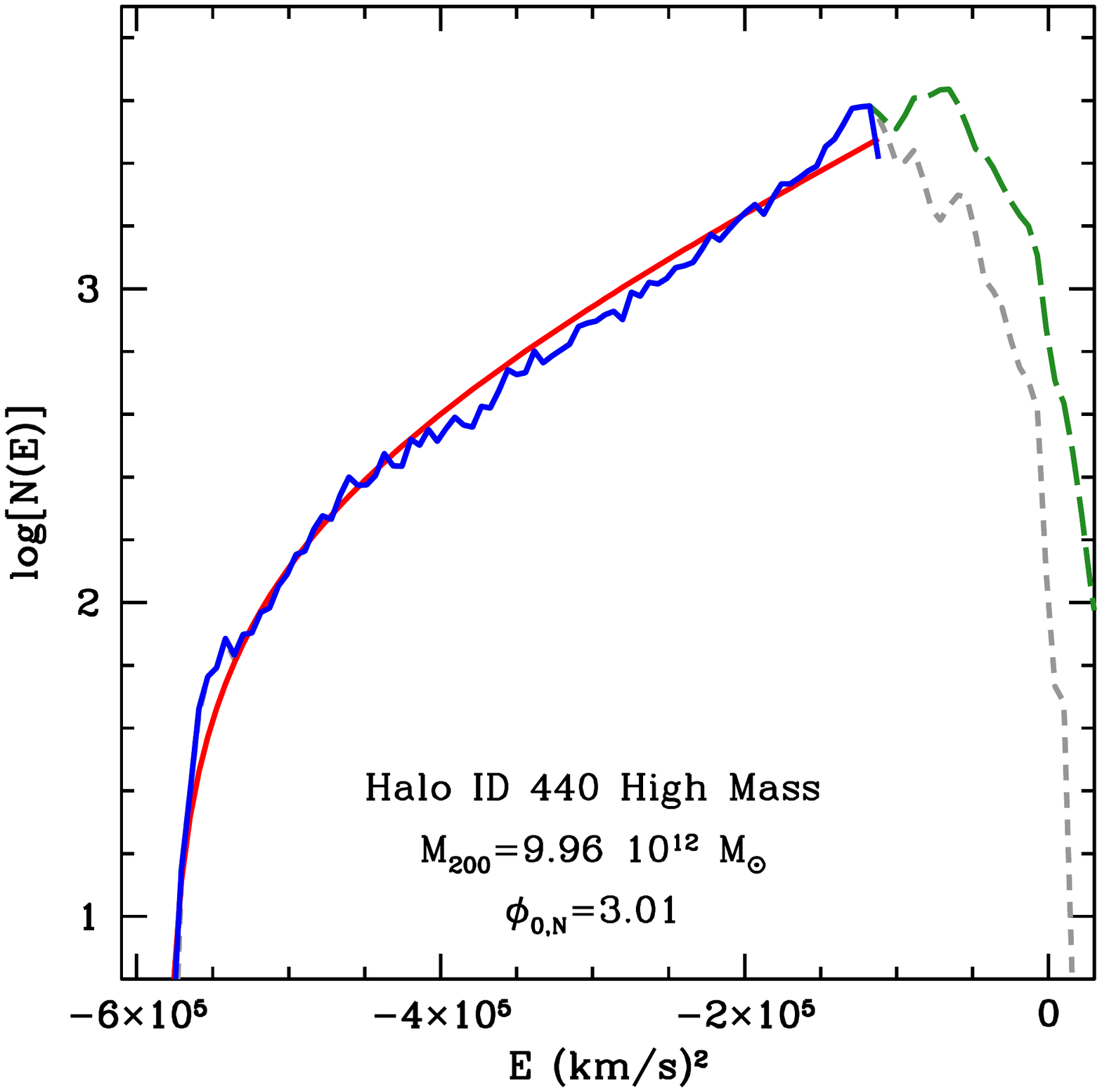}
\hfill
\includegraphics[width=.42\textwidth]{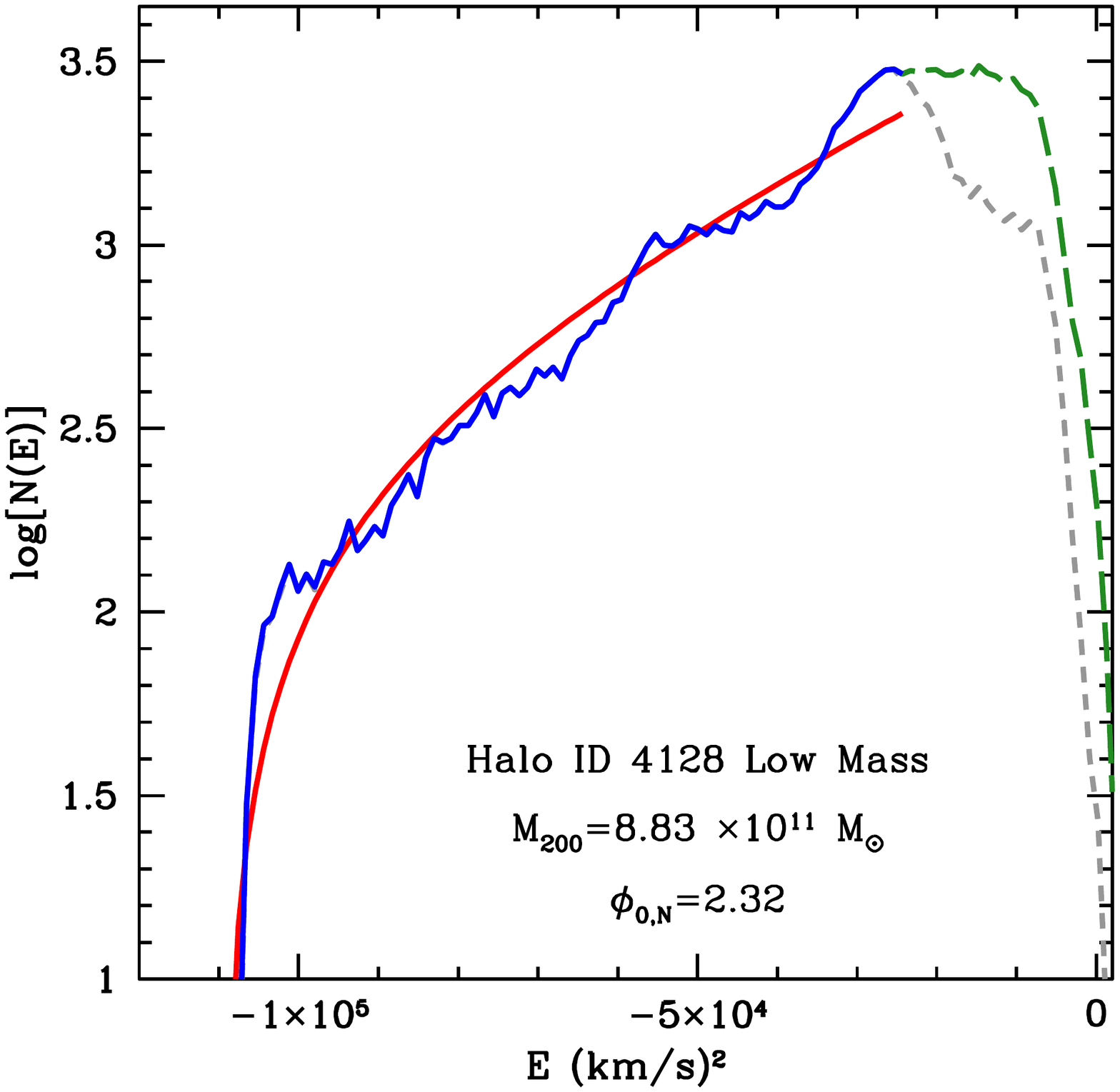}
\vspace{+25pt}
\caption{DARKexp fits to the differential energy distributions of selected high mass (left) and low mass (right) halos, presented in the same order as in Figure~\ref{fig:densityfits}. Top panels show halos that are fit very well, middle panels contain fits of `intermediate' quality, and bottom panels show `poor' fits. The virial mass and fitted $\phi_{0,N}$ are shown in each panel. Solid jagged (blue) lines show particle energy distributions truncated at $E=\Phi(r_{200})$, while the smooth solid (red) curve shows the DARKexp fit. Short-dash (gray) lines show $N(E)$ of particles within $r_{200}$, and long-dash (green) line shows that of all particles in the corresponding halo file.}
\label{fig:energyfits}
\end{figure}

Figure~\ref{fig:energyfits} shows a few examples of DARKexp $N(E)$ fits to selected Millennium-II halos. The top, middle, and bottom rows show examples of very good, intermediate quality, and poor fits. (Figure~\ref{fig:densityfits} plots density and density slope profiles fits to the same six systems.)  Solid jagged (blue) lines show Millennium-II particles energy distributions truncated at $E=\Phi(r_{200})$, while the smooth solid (red) curve shows the DARKexp fit. The fitted value of $\phi_0$, $\phi_{0,N}$ is shown in each panel. Short-dash (gray) lines show the energy distribution of particles within $r_{200}$, and long-dash (green) lines show that of all particles in the corresponding halo file. The fact that the $N(E)$ peak of the short-dash line (representing particles with $r\leq r_{200}$) coincides with $E=\Phi(r_{200})$ quite well for most halos implies that our choice of $E_{\rm hi}$ is sensible.

Note that the halos that are well fit by DARKexp follow it rather well, while those poorly fit by DARKexp tend to have more jagged $N(E)$ distributions, and are unlikely to be better fit by any smooth function of a few parameters.

\subsection{Average Halos}\label{avehaloN}

\begin{figure}
\includegraphics[width=.5\textwidth]{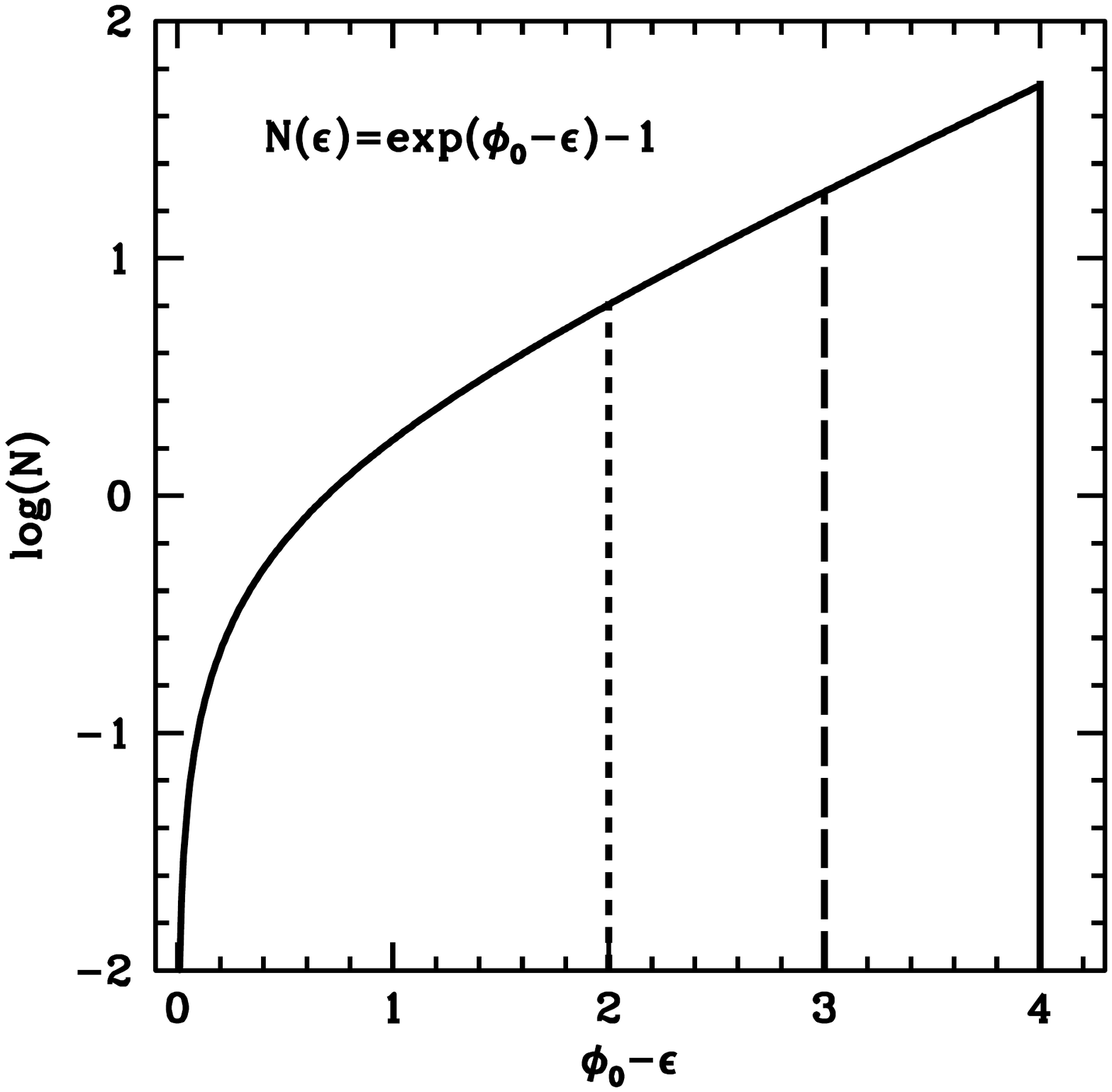}
\vspace{-75pt}
\includegraphics[width=.5\textwidth]{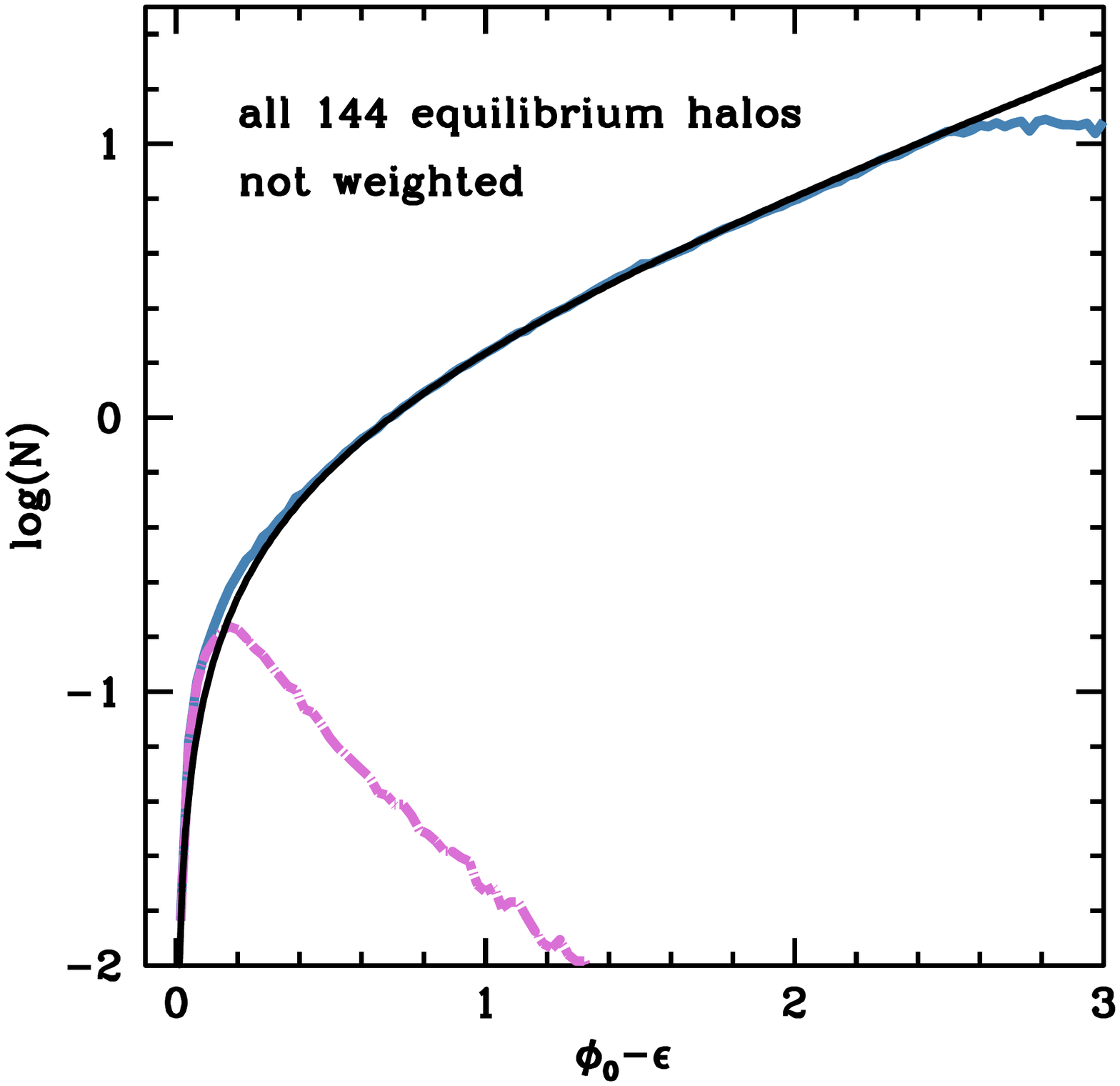}
\vspace{-75pt}
\includegraphics[width=.5\textwidth]{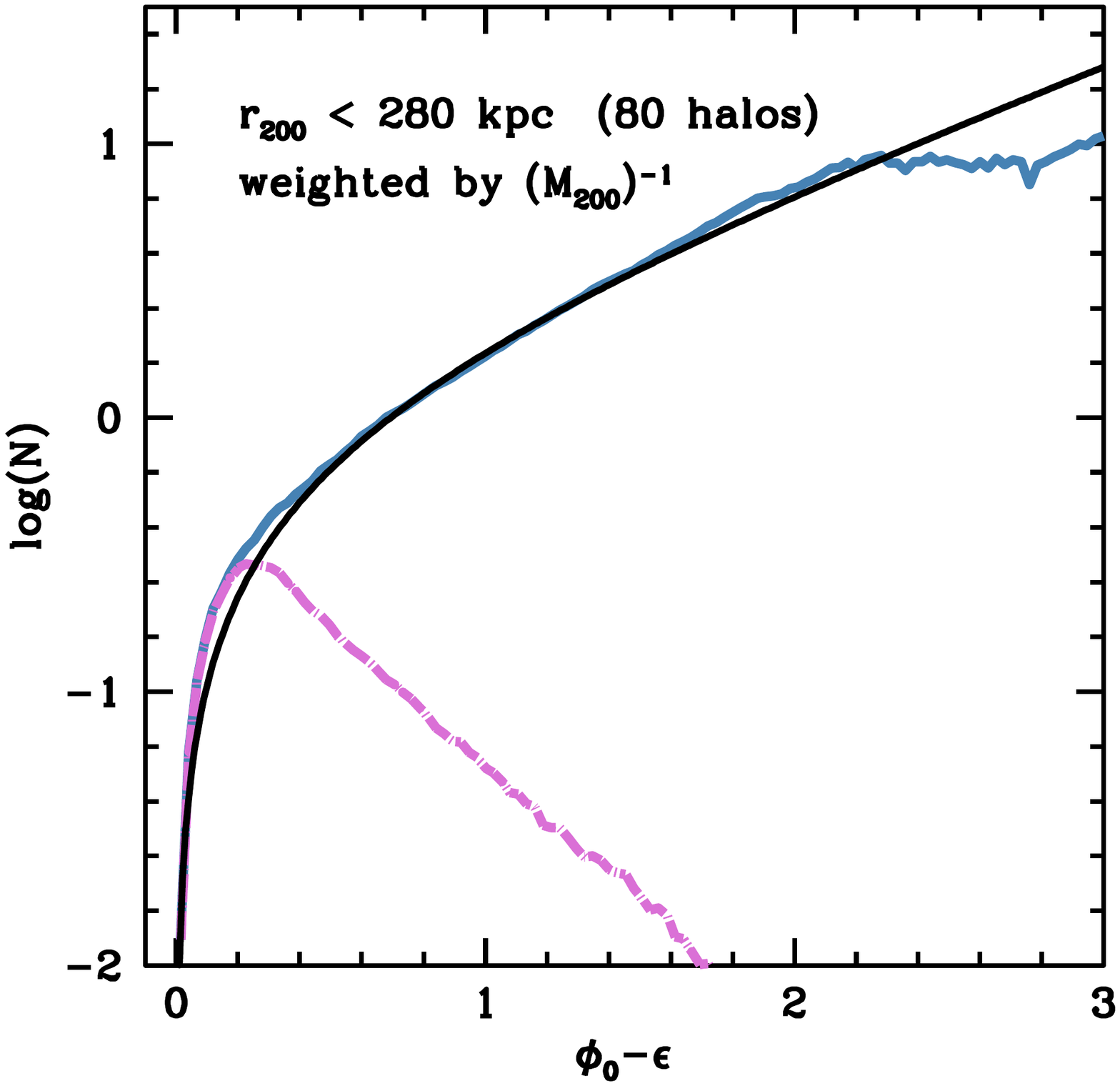}
\hfill
\includegraphics[width=.5\textwidth]{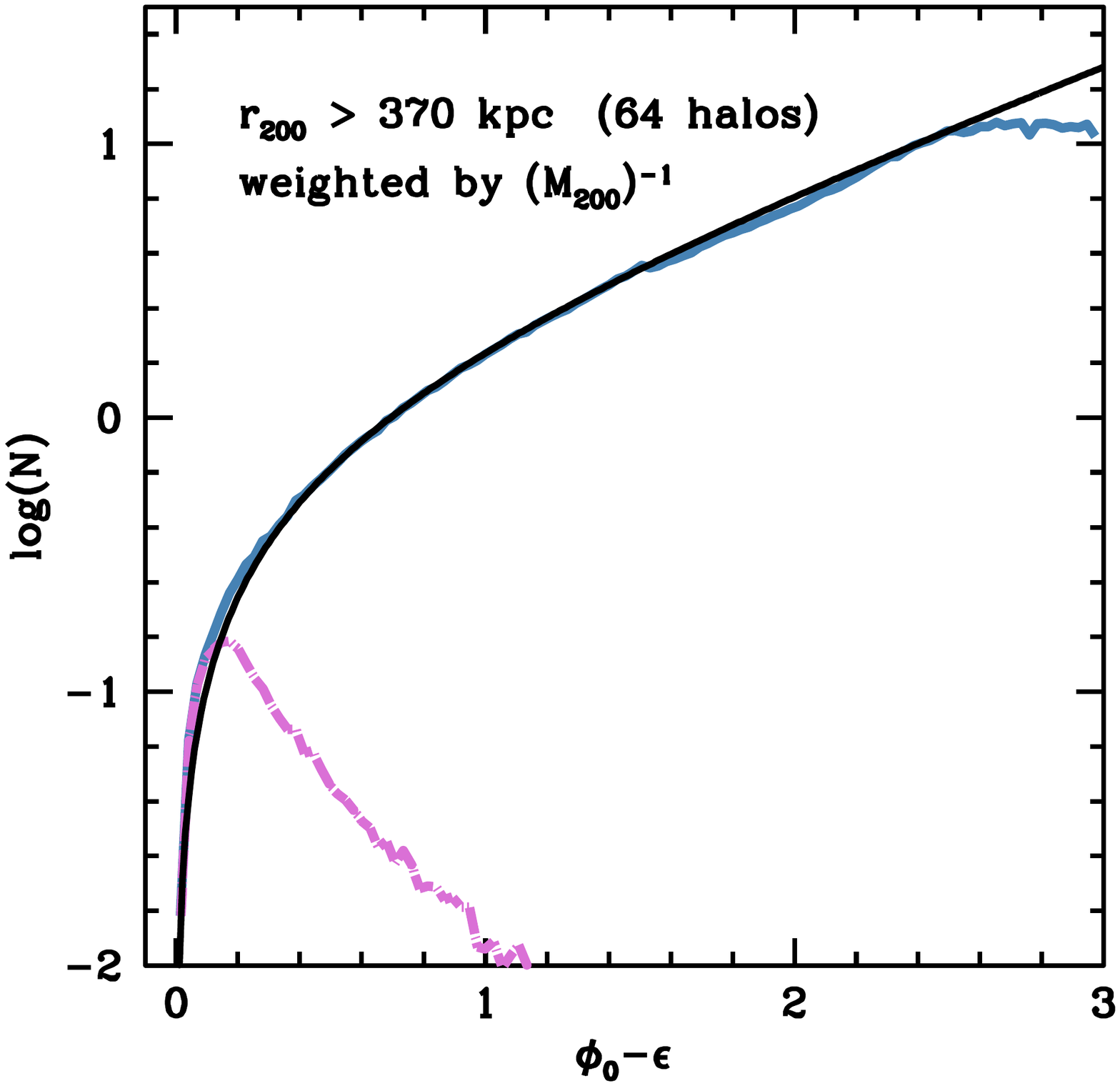}
\vspace{10pt}
\caption{{\it Upper left:} Examples of DARKexp differential energy distributions. When plotted using dimensionless $(\phi_0-\epsilon)$, all DARKexp differential energy distributions have exactly the same shape in the most bound energy range. The solid, long-dash and short-dash lines show DARKexp of $\phi_0=4, 3$, and $2$, respectively. {\it Upper right:} Average $N(\epsilon)$ of all 144 equilibrium Millennium-II halos, not weighted by the halo' virial mass (blue curves). Dot-dash magenta curve represents particles within $4 h^{-1}$~kpc (i.e. 4 smoothing lengths) from the halo centers. The black curve is DARKexp. {\it Lower panels:} Average $N(\epsilon)$ for halos with virial radii $r_{200}<280$ kpc (left) and $r_{200}>370$ kpc (right). In this case each halo's contribution was weighted by the inverse of its virial mass. Note that DARKexp fits the simulation results very well. High mass halos follow DARKexp better at bound energies than do the low mass halos, most likely because the resolution effects are less important in the former.}
\label{fig:sumNE}
\end{figure}

Because substructure and other density perturbations contribute to the fluctuations in the $N(E)$ distribution, we also present $N(E)$ distributions averaged over many halos. It is easy to obtain an average of many halos because of the following property of DARKexp. The shape of DARKexp differential energy distribution $N(\epsilon)$ plotted as a function of dimensionless energy, $\epsilon=\beta\,E$, is an exponential with a cutoff, eq.~\eqref{DARKexpEnergy}, and this basic shape is independent of the value of $\phi_0$, as is illustrated in the upper left panel of Figure~\ref{fig:sumNE}, where we use $(\phi_0-\epsilon)$ as the horizontal axis. The value of $\phi_0$ merely determines how far (to the right) along the horizontal axis the exponential part extends; the shorter the extent, the smaller the $\phi_0$. The shape of the more bound energy side of the $N(\epsilon)$ distribution, near $\phi_0-\epsilon\sim 0$ is the same for DARKexp of all $\phi_0$. 

This means that to get an average, we simply rescale the energies of the particles in each halo to dimensionless $\epsilon=\beta E$, shift them by $\phi_{0,N}$ to get $(\phi_0-\epsilon)$, and add the halos. The upper right panel of Figure~\ref{fig:sumNE} shows the average of all 144 equilibrium halos (solid light blue curve), where each particle of each halo contributed equally to the average $N(\epsilon)$. In the lower panels we show the results for the two mass ranges separately. Here we weighted each halo's contribution to the vertical axis equally, i.e. by the inverse of its virial mass. As seen in all three panels DARKexp provides an excellent fit to Millennium-II halos. Aside from the loosely bound end, $(\phi_0-\epsilon)\simgt 2$, where halos with small $\phi_0$'s do not contribute to the average, and friends-of-friends algorithm may have trouble separating out halo particles, the largest deviations from DARKexp are seen at very bound energies, $(\phi_0-\epsilon)\simlt 0.7$. (Even though the deviations of the average are confined to $(\phi_0-\epsilon)\simlt 0.4-0.5$, individual halos show deviations that extend to $(\phi_0-\epsilon)\sim 0.7$.)

At these energies, the low mass halos, with $r_{200}<280$~kpc, show a larger deviation than higher mass, $r_{200}>370$~kpc halos. These deviations---bumps---are likely due to numerical resolution effects. The dot-dash magenta lines show $N(\epsilon)$ only of the particles found within 4 smoothing lengths, or $4 h^{-1}$~kpc of the halo center. Their relative contribution to $N(\epsilon)$ appears to track the deviations from DARKexp at bound energies, and is expected to be larger for smaller mass halos because a fixed smoothing length encloses a larger fraction of a smaller halo's mass than a larger one's. The sense of the deviation also agrees with this interpretation: resolution effects artificially boosts particles' energies pushing them towards the less bound (or, right-ward) side of the distributions in Figure~\ref{fig:sumNE}. The corresponding effect in density profiles is to flatten out the central density cusp, which is seen in Figure~\ref{fig:profdev}, and described in Section~\ref{avehaloD}. Thus both, differential energy distributions and density profiles show effects of numerical resolution at the most bound energies and smallest radii. 

The bump described above is a systematic deviation of the energy distribution of Millennium-II halos from that of DARKexp. We would like to quantify these deviations. RMS is often used in such situations, but RMS is not able to differentiate between noise-like deviations and systematic ones. Therefore to separate out the influence of the bump we introduce a modified Lenz-Ising parameter,
\begin{equation}
S_{\rm{ising}}=\sum_{i=1}^{n-1}\,\Delta_i\,\Delta_{i+1},
\end{equation}
where $\Delta_i=N(\epsilon_i)_{\rm MII}-N(\epsilon_i)_{\rm DARKexp}$ is the difference between data and model at dimensionless energy $\epsilon_i$. In the standard Lenz-Ising model, $\Delta_i$ would be just the sign of $[N(\epsilon_i)_{\rm MII}-N(\epsilon_i)_{\rm DARKexp}]$ \citep{lee12}. We use the actual difference because of the additional information contained therein. For a model that fits well, i.e. when the data points are equally likely to be above or below the model, the expectation value for $S_{\rm{ising}}$ is zero. On the other hand, systematic deviations will lead to positive values of $S_{\rm{ising}}$. The absolute values of our modified $S_{\rm{ising}}$ parameter cannot be compared to some fiducial model, but they can be compared to each other. $S_{\rm{ising}}$ does depend on the binning, but we use very nearly the same binning for all halos. 

\begin{figure}
\includegraphics[width=.5\textwidth]{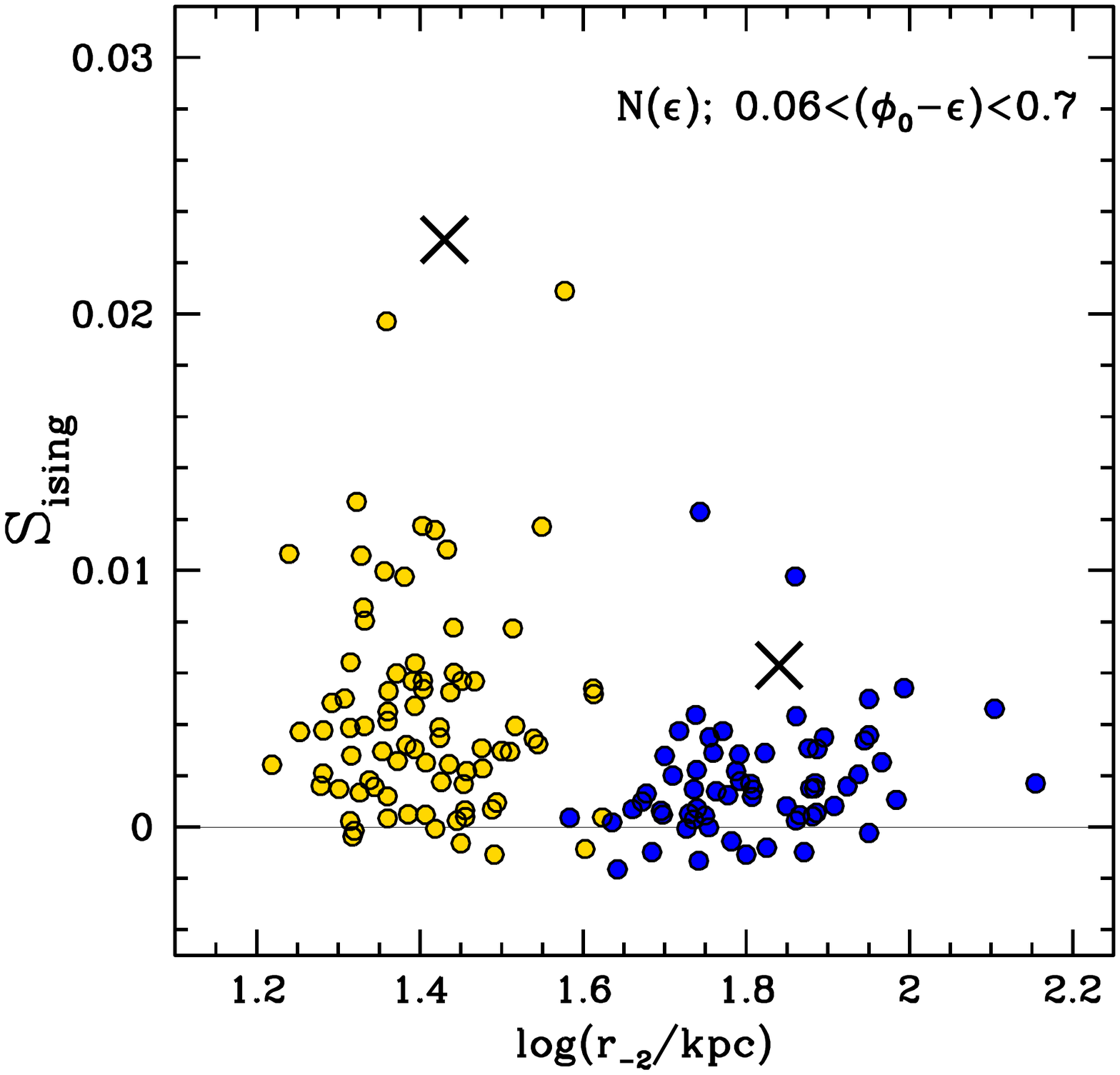}
\hfill
\includegraphics[width=.5\textwidth]{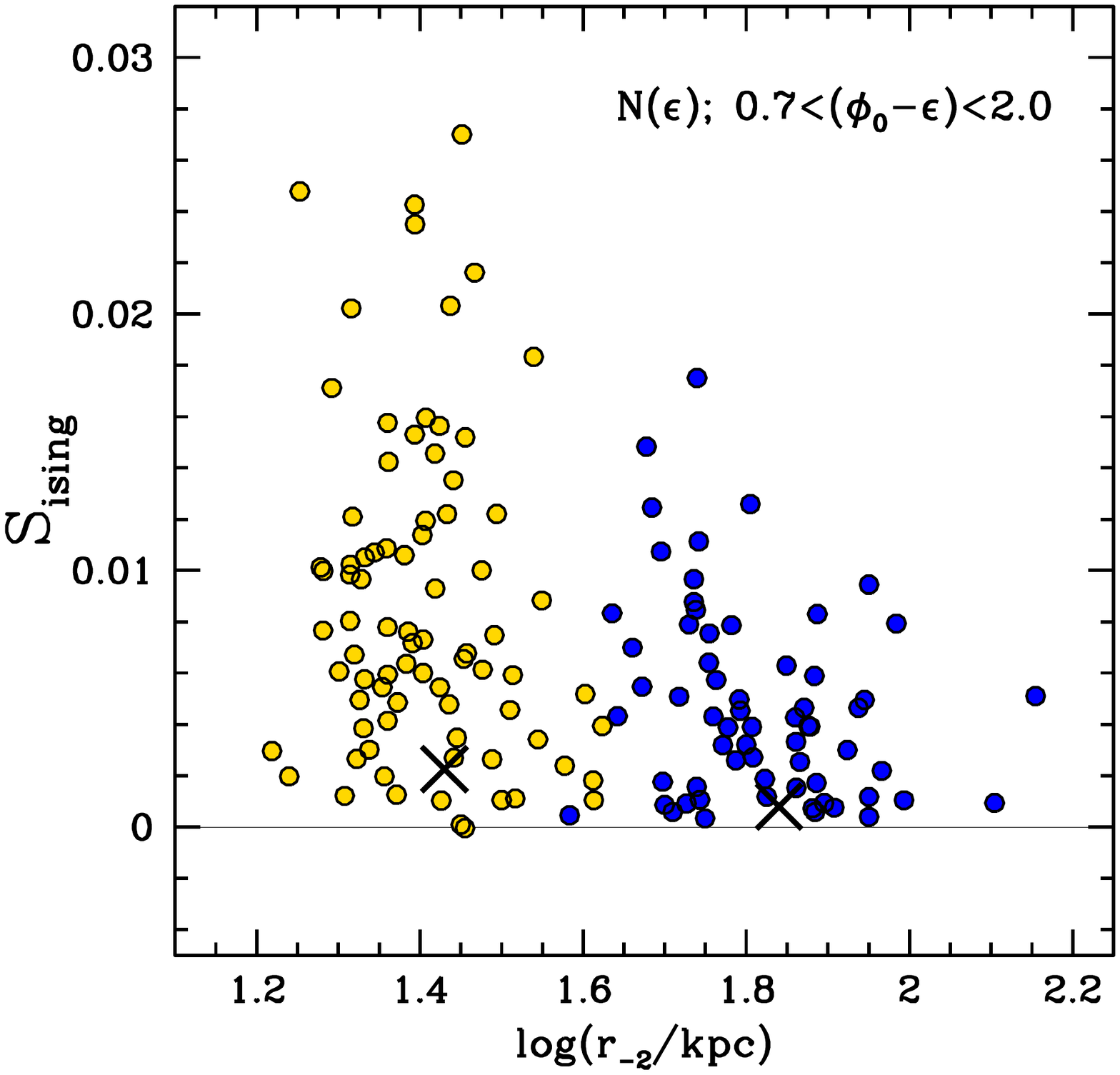}
\vspace{-60pt}
\caption{The (modified) ising model measure of how well DARKexp $N(\epsilon)$ fits 144 equilibrium Millennium-II halos vs. the halos' $r_{-2}$ radius. Low mass halos, $r_{200}<280$~kpc, are denoted by yellow points, and high mass halos, $r_{200}>370$~kpc, by blue points. The left panel is based on the very bound end of $N(\epsilon)$, where $0.06<(\phi_0-\epsilon)<0.7$, while the right panel is based on the loosely bound end of $N(\epsilon)$, where $0.7<(\phi_0-\epsilon)<2.0$. The large crosses show $S_{\rm{ising}}$ for average $N(\epsilon)$ distributions shown in Figure~\ref{fig:sumNE}.
}
\label{fig:ising}
\end{figure}

We calculate $S_{\rm{ising}}$ for high and low mass halos, and for two segments of $N(\epsilon)$ separately, and plot them in Figure~\ref{fig:ising}. The left and right panels show $S_{\rm{ising}}$ plotted against halos' $r_{-2}$ for the portion of $N(\epsilon)$ where $0.06<(\phi_0-\epsilon)<0.7$, and $0.7<(\phi_0-\epsilon)<2.0$, respectively. The values 0.06 and 0.7 were chosen to isolate the bump based on the upper right and the two lower panels of Figure~\ref{fig:sumNE}. The value 2.0 corresponds roughly to the lowest $\phi_{0,N}$ value; at $(\phi_0-\epsilon)\simgt 2$ not all halos contribute to $N(\epsilon)$, and the average Millennium-II differential energy distribution begins to deviate strongly from that of DARKexp. The circular points in Figure~\ref{fig:ising} show 144 equilibrium halos, with light yellow (dark blue) points representing low (high) mass systems. A number of conclusions can be drawn from the plot. First, nearly all values of $S_{\rm{ising}}$ are above zero, meaning that the bulk of the halos' $N(\epsilon)$ deviate from that of DARKexp over extended $\epsilon$ intervals. High mass halos deviate from DARKexp less than low mass halos; this is supported by the KS test comparing the two distributions in each of the two panels. 

There are marked difference between the deviations from DARKexp at bound energies, $0.06<(\phi_0-\epsilon)<0.7$, and loosely bound energies, $0.7<(\phi_0-\epsilon)>2.0$, and between high mass and low mass systems. Crosses (two per panel in Figure~\ref{fig:ising}) show $S_{\rm{ising}}$ for the average $N(\epsilon)$ distributions, as measured from the corresponding two lower panels of Figure~\ref{fig:sumNE}. At very bound energies (left panel of Figure~\ref{fig:ising}) the two averages over halos (crosses) show deviations that are significantly larger than those of individual halos. This means that the deviations are common to most halos, and while the random halo-to-halo noise is reduced by the averaging, the systematic deviation adds up. It is also worth noting that the massive halos, $r_{200}>370$~kpc, have a smaller deviation (bump) compared to the less massive ones, $r_{200}<280$~kpc, i.e. the cross corresponding to the former is closer to $S_{\rm{ising}}=0$.

At the loosely bound end of $N(\epsilon)$ (right panel of Figure~\ref{fig:ising}) the two averages over low and high mass halos (crosses) have $S_{\rm{ising}}$ that are significantly lower, and are nearly zero compared to those of individual halos. This means that at these energies the deviations of Millennium-II $N(\epsilon)$ from that of DARKexp are random halo-to-halo variations, likely driven by substructure and other density perturbations, and therefore nearly disappear upon averaging of halos. To sum up, Figure~\ref{fig:ising} quantifies what can be deduced by looking at Figures~\ref{fig:energyfits} and \ref{fig:sumNE}.

\section{\boldmath{Comparing ${\phi_0}$ values from ${\rho(r)}$ and ${N(E)}$ fitting}}\label{comp}

For an isolated relaxed halo, the DARKexp shape parameters obtained from two independent types of fitting, to the density profile, $\phi_{0,D}$, and to the differential energy distribution, $\phi_{0,N}$, should agree. However, Millennium-II simulated halos are embedded in a larger cosmological setting, with other nearby halos and unbound particles. This makes it hard to separate out individual halos, and no criteria or algorithm, for example friends-of-friends, can do so perfectly. Even two different methods that work in configuration space will pick out somewhat different sets of particles in the outskirts a halo, and will not perfectly agree on halo membership, and hence on the values of the fitted parameters. The two methods we use to decide which particles belong to a halo work in different spaces, in configuration and energy spaces, respectively, and so are not expected to agree perfectly. The differences in halo particle membership lead to differences in fitted shape parameters, as shown below.

Figure~\ref{fig:NEone1} plots $\phi_{0,N}$ vs. $\phi_{0,D}$ for all 200 halos. The dark blue (light yellow) points represent systems with $r_{200}>370$~kpc ($r_{200}<280$~kpc). Filled symbols represent systems in equilibrium, while empty symbols those that did not pass one or both of the criteria described in Section~\ref{deteq}. The inset in the lower right shows the histogram distribution of $\phi_{0,D}$ values of equilibrium systems only, using the same color scheme as the points in the main plot. The inset in the upper left shows the distribution of $\phi_{0,N}$ values. Two trends are seen: (1) $\phi_{0,N}$'s are systematically lower than $\phi_{0,D}$'s for both halo mass ranges; (2) $\phi_{0,N}$'s of low mass systems are systematically below those of the high mass systems.

\begin{figure}
\centering
\includegraphics[width=.85\textwidth]{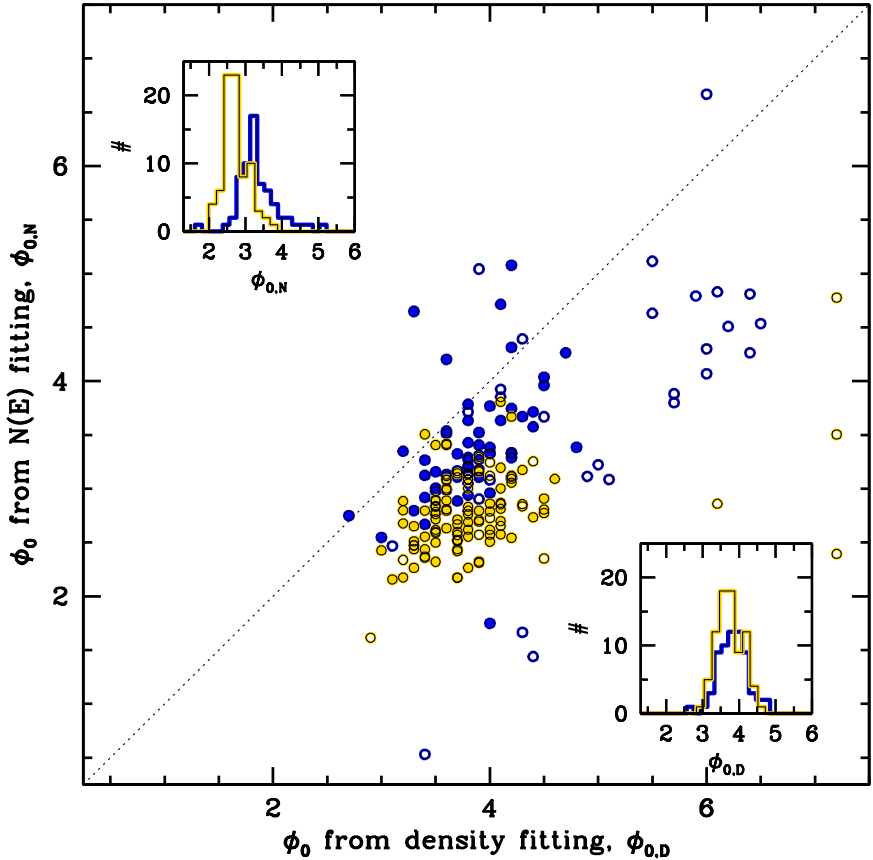}
\vspace{-80pt}
\caption{DARKexp shape parameter $\phi_0$, from independent fits to density profiles, $\phi_{0,D}$, and differential energy distributions, $\phi_{0,N}$ of Millennium-II halos. The blue (yellow) points represent halos with $r_{200}>370$~kpc ($r_{200}<280$~kpc). Filled (empty) circles represent halos that did (did not) pass the equilibrium halo criteria of Section~\ref{deteq}. The insets show histograms of distribution of $\phi_0$ (upper left: $\phi_{0,N}$, lower right: $\phi_{0,D}$) of equilibrium halos of low (yellow lines) and high (blue lines) mass.}
\label{fig:NEone1}
\end{figure}

\begin{figure}
\centering
\includegraphics[width=.85\textwidth]{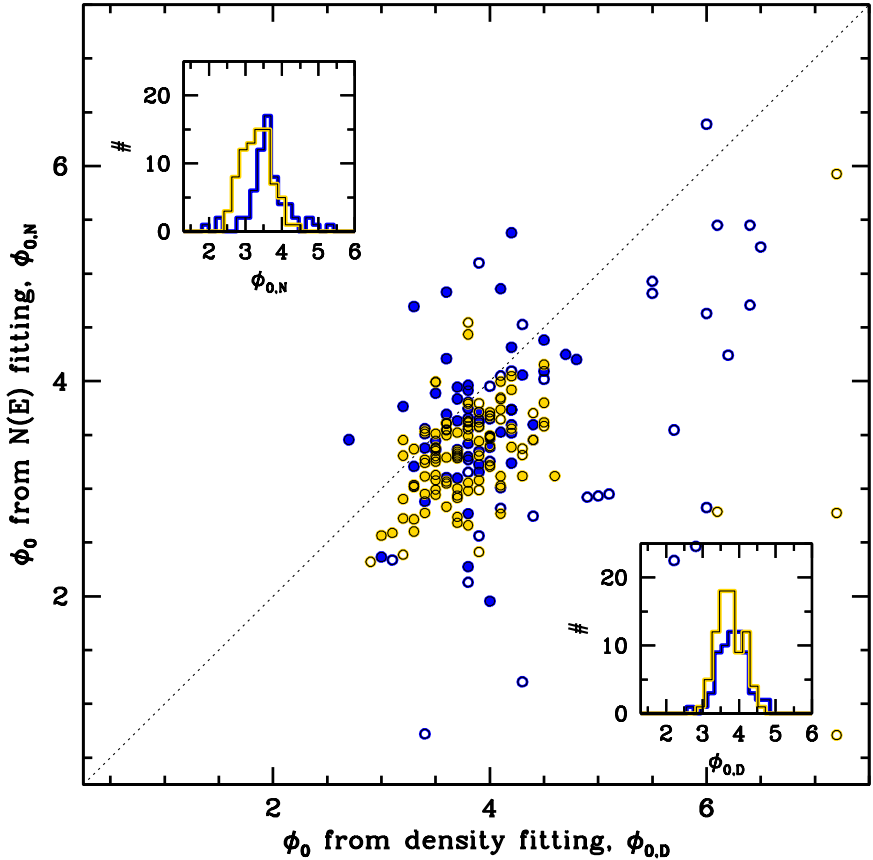}
\vspace{-80pt}
\caption{Similar to Figure~\ref{fig:NEone1}, but here the fits to the differential energy distribution of Millennium-II halos were done excluding a region with dimensionless energies $0.06<(\phi_0-\epsilon)<0.7$.}
\label{fig:NEone2}
\end{figure}

Let us first concentrate on (2). The bump in $N(E)$ described in Section~\ref{avehaloN} is more significant for low mass systems. When fitting DARKexp to Millennium-II halos' $N(E)$ this bump will tend to lower $\phi_{0,N}$ values because its presence reduces the vertical extent of the exponential portion of DARKexp $N(E)$. We can check if that is the case by excluding the bump region while fitting. As already discussed in Section~\ref{avehaloN}, lower panels of Figure~\ref{fig:sumNE} show that this region is approximately confined to $0.06<(\phi_0-\epsilon)<0.7$. Excluding this region from the fit produces results shown in Figure~\ref{fig:NEone2}. Values of $\phi_{0,N}$ of low mass halos have gone up by, typically, $\sim 0.65$, while those of high mass halos, by $\sim 0.35$, which resolves most of the issue (2) and some of (1). 

Let us now return to the remaining (1), i.e., the reason why $\phi_{0,N}$ values of both mass ranges are systematically lower than the corresponding $\phi_{0,D}$ values, by $\delta\phi_{0,N}\sim 0.3$ (Figure~\ref{fig:NEone2}). Because there is no reason to suspect that the density fits would give biased $\phi_{0,D}$ values, we suspect that the fault lies with fitted $\phi_{0,N}$ values. Recall that the Millennium-II $N(E)$ were constructed from particles with total energies below $\Phi(r_{200})$ (Section~\ref{indivhalos}), which missed some bound particles whose energies are larger than that. Figure~\ref{fig:energyfits} shows that $N(E)$ distributions of particles found within $r_{200}$ (short-dash gray lines) tend to extend to energies less bound (less negative) than $\Phi(r_{200})$ and therefore particles with $E\leq\Phi(r_{200})$ are a subset of particles with $r\leq r_{200}$. Furthermore, bound particles with $r\leq r_{200}$ are a subset of all particles that are bound to the halo, because at any given time, some particles whose pericenters are within $r_{200}$ will find themselves outside $r_{200}$, on the apocenter portion of their elliptical orbits. If these were included they would have extended the exponential part of $N(E)$ to less bound energies and increased the fitted values of $\phi_{0,N}$. 

If our estimated $\delta\phi_{0,N}$ is to be attributed to these particles, one can calculate the fraction of halo mass that has energies larger than $\Phi(r_{200})$. The total mass contained in a halo can be obtained by integrating eq.~\ref{DARKexpEnergy}: $M(\phi_0)=A[\exp(\phi_0)-(\phi_0+1)]$. Adopting typical $\phi_{0,D}\approx 3.75$, and typical $\phi_{0,N}\approx 3.45$ (see Figure~\ref{fig:NEone2}) implies that $[M(\phi_{0,D})-M(\phi_{0,N})]/M(\phi_{0,D})\approx 28$\% of halo bound particles have energies larger than $\Phi(r_{200})$. 

\section{Conclusions}

We show that the theoretically derived DARKexp model for relaxed collisionless self-gravitating systems provides very good fits to energy and density distributions of individual Millennium-II pure dark matter halos over a mass range spanning a factor of $\sim 50$. This is the first study that presents DARKexp fits to both $N(E)$ and $\rho(r)$, and compares the two. 

To reduce the noise in the energy and density distributions of individual halos, and to reveal the systematic deviations between Millennium-II halos and DARKexp we calculate the average $\rho(r)$ and $N(\epsilon)$ for high and low mass systems separately. The average density profile shows that both mass ranges behave similarly at large radii, while at small radii, interior to 3 numerical smoothing lengths, they show features consistent with resolution effects. The radial location of these features in high and low mass halos further supports this interpretation. If resolution effects were negligible, DARKexp would likely provide an excellent fit to the average Millennium-II halo density profile.

The equation for DARKexp $N(\epsilon)$, where $\epsilon$ is dimensionless energy, is such that it allows many halos to be added together to obtain an average. The average shows that DARKexp provides an excellent fit to the less bound `half' of $N(\epsilon)$, between $0.7<(\phi_0-\epsilon)<2.0$, and that the deviations from DARKexp in individual halos are random, and are likely driven by substructure and other density perturbations. At bound energies, $(\phi_0-\epsilon)<0.7$, there is a small, bump-like deviation of Millennium-II halos from DARKexp, which is more pronounced in low mass systems ($r_{200}<280$~kpc), and almost absent in high mass systems ($r_{200}>370$~kpc); see lower panels of Figure~\ref{fig:sumNE}. This bump is a systematic deviation and is most likely due to the effects of numerical resolution, which artificially boost the energies of the most bound particles. Its presence and relative strength in high and low mass systems is consistent with the deviations of the Millennium-II average density profiles from those of DARKexp.

After excluding the affected energy range from the fits, the values of $\phi_0$ obtained from density profile fitting, $\phi_{0,D}$, correlate with those from $N(E)$ fitting, $\phi_{0,N}$, but are offset from a 1-to-1 relation by $\phi_{0,D}-\phi_{0,N}\approx 0.3$. This offset, which is comparable to the scatter in the determination of $\phi_{0,D}$, is related to the fact that a halo boundary in energy space and in configuration space are only loosely related, and selecting particles with total energies less than $\Phi(r_{200})$, as we have done for $N(E)$ fitting, excludes many particles that are bound to the halo. A virial radius, $r_{200}$, boundary in configuration space is more inclusive, but it still excludes particles that happen to be closer to their apocenters at the end of the simulation. 

The upper right and two lower panels of Figure~\ref{fig:sumNE}, which show DARKexp model fits to the differential energy distribution of Millennium-II halos, are the main results of this paper. The DARKexp $N(\epsilon)$, which is an exponential with a cutoff, eq.~\ref{DARKexpEnergy}, has no shape parameters; $\phi_0$ only sets the extent, on the energy axis, of the exponential portion of $N(\epsilon)$. The very good agreement between the differential energy distribution found in simulated dark matter halos and modeled by DARKexp, provide strong support for using DARKexp as a realistic model of the energy and density distributions of dark matter halos.

\section*{Acknowledgements}

This material is based upon work supported by the National Science Foundation Graduate Research Fellowship under Grant No. 00039202.

\bibliographystyle{unsrt}
\bibliography{bibfile}

\end{document}